\documentclass[12pt]{article}
\usepackage{tikz}
\usepackage[font=small,labelfont=bf]{caption}
\usetikzlibrary{mindmap,backgrounds}
\usetikzlibrary{positioning}
\usepackage{longtable}
\usepackage{booktabs}
\usepackage{makeidx}
\usepackage{algorithm}
\usepackage{xcolor}
\usepackage{algpseudocode}
\usepackage{algcompatible}
\usepackage{float}
\usepackage{times}
\usepackage{amssymb}
\usepackage{amsmath}
\numberwithin{equation}{section}
\usepackage{setspace}
\usepackage{hyperref}
\usepackage{xurl}
\usepackage{rotating}
\usepackage{color}
\usepackage{graphicx}
\usepackage{rotating}
\usepackage{pdflscape}
\usepackage{epstopdf}
\usepackage[round,authoryear,comma]{natbib}
\usepackage{natbib,hyperref}
\usepackage{booktabs}
\usepackage{setspace}
\usepackage{pgfplots}
\usepackage{subcaption}
\usepackage{xcolor}
\usepackage{multirow}
\usepackage{array}
\usepackage{placeins}
\usepackage{actuarialangle}
\usepackage{tabularx}
\usepackage{booktabs}
\usepackage{tikz}
\usepackage[framemethod=TikZ]{mdframed}
\usepackage{lineno, blindtext}
\usepackage{natbib}
\usepackage{bbm}
\usetikzlibrary{matrix,shapes,chains,positioning,decorations.pathreplacing,arrows}
\hypersetup{
	colorlinks   = true, %Colours links instead of ugly boxes
	urlcolor     = blue, %Colour for external hyperlinks
	linkcolor    = blue, %Colour of internal links
	citecolor   = blue %Colour of citations
}

\newlength{\Oldarrayrulewidth}

\usepackage{hyperref} % dixon
\providecommand{\U}[1]{\protect\rule{.1in}{.1in}}
\parskip=\medskipamount

\newcolumntype{C}[1]{>{\centering\arraybackslash}m{#1}}

\begin{document}

%\title{Graph Neural Networks for Pricing Catastrophe Derivative Contracts}
% \title{CATNet: A geometric deep learning approach for CAT bond spread prediction in the primary market}
% \author{Dixon Domfeh\footnote{Corresponding author}  \\ \small College of Computing, Georgia Institute of Technology, Atlanta, United States\\ \small \texttt{dnkwantabisa3@gatech.edu} \and 
% Saeid Safarveisi \\ \small Actuarial Research Group, AFI, Faculty of Economics and Business,
% KU Leuven, Leuven, Belgium \\ \small \texttt{saeid.safarveisi@kuleuven.be}
% }
% \date{Version: \today }

\title{CATNet: A geometric deep learning approach for CAT bond spread prediction in the primary market}

\author{
  Dixon Domfeh\thanks{Corresponding author. College of Computing, Georgia Institute of Technology, Atlanta, United States. \texttt{dnkwantabisa3@gatech.edu}}
  \and
  Saeid Safarveisi\thanks{Actuarial Research Group, AFI, Faculty of Economics and Business, KU Leuven, Leuven, Belgium. \texttt{saeid.safarveisi@kuleuven.be}}
}

\maketitle
\hrule
\begin{abstract}
\noindent Traditional models for pricing catastrophe (CAT) bonds struggle to capture the complex, relational data inherent in these instruments. This paper introduces CATNet, a novel framework that applies a geometric deep learning architecture, the Relational Graph Convolutional Network (R-GCN), to model the CAT bond primary market as a graph, leveraging its underlying network structure for spread prediction. Our analysis reveals that the CAT bond market exhibits the characteristics of a scale-free network, a structure dominated by a few highly connected and influential hubs. CATNet demonstrates higher predictive performance, significantly outperforming strong Random Forest and XGBoost benchmarks. Interpretability analysis confirms that the network's topological properties are not mere statistical artifacts; they are quantitative proxies for long-held industry intuition regarding issuer reputation, underwriter influence, and peril concentration. This research provides evidence that network connectivity is a key determinant of price, offering a new paradigm for risk assessment and proving that graph-based models can deliver both state-of-the-art accuracy and deeper, quantifiable market insights.  \\

\smallskip
%\noindent \textbf{Keywords}: %
% \noindent \textbf{JEL classification:} G12; G22; G01; C45; D85; C55; Q54


\end{abstract}
\hrule

\newpage
\section{Introduction}
% Catastrophe (CAT) bonds are financial instruments that transfer risk from insurers to investors, covering natural disasters. Essentially, CAT bonds allow insurance companies to pass on the risk of natural disasters covered by their policies to investors for a price. The yields on CAT bond contracts reflect their exposure to both climate and interest rate risks.

Catastrophe (CAT) bonds are financial instruments that transfer risk related to natural disasters from insurers to investors for a price. Modeling these instruments is crucial for pricing, risk assessment, and portfolio management. 
Traditional econometrics and machine learning (ML) approaches have been applied to CAT bond transaction datasets. However, they face significant challenges. Many machine learning models assume data points are independent and identically distributed (IID), a condition often violated in CAT bond datasets due to temporal, spatial, and peril-based dependencies. Market conditions and catastrophic events can influence multiple contracts simultaneously \citep{A1}, creating time-based correlations, while bonds covering the same geographical regions can lead to spatial correlations \citep{A2}.

Another challenge is the high cardinality of categorical variables in the data. To manage this, researchers often resort to data manipulation, such as combining perils or grouping regions, (see example, \cite{A4}, \cite{A3}). While this reduces dimensionality, it comes at the cost of losing granular information that may be crucial for accurate modeling. These limitations hinder the ability of traditional models to effectively represent the complex relationships between the various entities involved in CAT bond contracts. Important relationships and patterns that exist at more detailed levels (e.g., specific states or perils) may be obscured or lost entirely, potentially leading to less accurate models. 

Traditional ML models like linear regression, decision trees, and some ensemble methods (like random forest and XG Boost) struggle with datasets containing categorical variables with many unique values. Each unique combination of categories (like state and peril) effectively becomes a new feature, drastically increasing the dimensionality of the dataset and making it difficult to train a robust model. Furthermore, traditional ML models struggle to represent the complex relationships between different entities involved in CAT bond contracts, such as issuers, perils, and regions. This limitation hinders their ability to accurately predict bond performance and assess risk.

Geometric Deep Learning (GDL), and specifically Graph Neural Networks (GNNs), offer a compelling solution. GNNs are designed to operate directly on graph structures, making them ideal for modeling the intricate web of relationships in the CAT bond market. Nodes in the graph can represent entities like contracts, perils, regions, and issuers, while edges capture relationships between them. GNNs can effectively process high-cardinality categorical variables by representing them as nodes and edges in a graph, thus capturing intricate relationships and preserving valuable granular information, as demonstrated in \citep{A6} for recommender systems. The message passing mechanism of GNNs enables the model to aggregate information from neighboring nodes \citep{A7}, capturing the influence of related contracts and perils. Through representation learning, GNNs learn embeddings that encapsulate both the features of nodes and their structural relationships, enabling effective generalization to unseen data \citep{A8}. GNNs have shown superior performance in domains with complex relational structures, such as social networks \citep{A9}, biological networks \citep{A10}, and recommendation systems \citep{A11}, suggesting their potential effectiveness for CAT bond modeling. 

In this paper, we introduce CATNet, a novel framework that applies a Relational Graph Convolutional Network (R-GCN) to this domain, making several key contributions. CATNet significantly outperforms state-of-the-art tabular models (Random Forest and XGBoost) in spread prediction using only raw bond contract data. This suggests that the market is largely efficient and that the primary predictive challenge lies in effectively representing data complexity.

% We demonstrate that CATNet significantly outperforms the state-of-the-art Random Forest (RF) and XG Boost tabular models in spread prediction using only the raw information within the bond contracts, which suggests the market is largely efficient and that the primary challenge lies in effectively representing data complexity. 

Our analysis also reveals that the CAT bond market has a scale-free network structure, a critical insight into its organization and potential systemic vulnerabilities. Furthermore, we provide interpretable results, showing how network centrality measures act as quantitative proxies for long-held industry intuition about issuer reputation and peril concentration. Ultimately, this work establishes a new paradigm for analyzing CAT bond primary market by prioritizing the learning of relational structure over traditional feature engineering and data manipulation.

We begin by introducing the data and our graph representation method in Sections \ref{sec2} and \ref{sec3}. In Section \ref{sec4}, we detail the R-GCN architecture and its application to risk premium prediction. We provide a comprehensive discussion on the results and model interpretation in Section \ref{sec5}, and finally conclude the paper in Section \ref{sec6}, highlighting areas for future research.

\subsection{A brief literature review}

Research on catastrophe bond pricing can be broadly categorized into two areas--valuation and prediction. Valuation focuses on modeling the underlying risk characteristics of CAT bonds, including factors such as catastrophe risk (claim amount and intensity of perils) and interest rate risk (e.g. \cite{Safarveisi2025}; \cite{Domfeh2024}; \cite{Ibrahim2022}). Prediction, on the other hand, aims to forecast the risk premium of a CAT bond contract based on existing contract characteristics, such as the cedent, underwriter, and maturity. Our paper focuses on this second area--the prediction of risk premiums.

Early research on CAT bond pricing primarily employed exploratory frameworks to identify variables that were both theoretically relevant and statistically significant in explaining CAT bond prices \citep{A3}. \cite{A30} developed the first model to characterize the behavior of the CAT bond market, a seminal work that paved the way for subsequent studies. Studies that followed built upon this foundation, investigating the impact of specific events and attributing CAT bond prices in both primary and secondary markets to various contract-specific and macroeconomic factors. For example, \cite{Ahrens2014} investigated how catastrophe risks are priced by examining the impact of Hurricane Katrina on CAT bond prices. Their analysis revealed that catastrophe risk prices are influenced by the underlying peril, the expected loss, the wider capital market cycle, and the risk profile of the transaction. \cite{A43} utilized a generalized additive model to examine the factors that affect CAT bond premiums and identified similar factors, including insurance underwriting cycles, rating class, issuer, catastrophe risk modeler, territory covered, and trigger type, as relevant drivers of CAT premium in the primary market. \cite{Braun2016} confirmed that the expected loss is the primary driver of prices, while also highlighting the importance of other factors.

While these early studies provided valuable insights into CAT bond pricing, \cite{Major2019} argued that they did not directly address the business need for spread prediction. However, these early research laid the foundation for the application of modern machine learning approaches in the CAT bond domain.  More recently, several studies have explored the application of machine learning for CAT bond price prediction. \cite{A4} compared linear regression, random forests, and neural networks for pricing CAT bonds. \cite{A3} proposed a random forest model to predict spreads in the primary market. \cite{A5} investigated the forecasting accuracy of random forests and neural networks for predicting CAT bond returns in the secondary market. \cite{Chen2024} introduced a probabilistic machine learning approach for pricing catastrophe bonds in the primary market. While these ML frameworks have advanced our understanding of CAT bond pricing, they still have limitations. These limitations include selection bias, predictor interactions, non-linearities \cite{A3}, and data complexity. Previous studies often excluded contracts with missing data or outliers, potentially leading to selection bias and loss of information. Additionally, traditional ML models often struggle to capture the complex interdependencies within CAT bond data.

\section{Data} \label{sec2}
 Our dataset comprises 803 catastrophe bond contracts issued in the primary market between 1999 and 2021, of which approximately 64\% cover multiple perils and 34\% are issued across multiple countries. Although most of these transactions have matured, the underlying interactions—between entities, peril types, and geographic regions—provide a rich dataset. This allows for the characterization of the catastrophe bond market as a dynamic network, revealing its structural evolution over time. Conventionally, CAT bond analysis combines contract data with exogenous variables (e.g., BBB corporate bond spreads, Guy Carpenter Index) to reflect the prevailing interest rate and insurance market environments. This study diverges from that precedent, operating on the premise that such external factors are already incorporated into the contract pricing. Therefore, we exploit the relationships within the contract data to demonstrate that we can accurately predict risk premiums without relying on these additional features. In our data preprocessing, we removed the ``Expected Excess Return'' variable due to its high correlation (approx. 90\%) with our target variable, the ``Risk Premium over LIBOR.'' Because excess return is a direct function of the risk premium and expected loss, it offers no additional explanatory power for our model.\\

The data were hand-collected from two primary sources: Lane Financial reports and the Artemis Deal Directory. Lane Financial reports provided quantitative contract details, such as risk premium, probability of loss, expected loss, issue amount, and year of issuance. The Artemis Deal Directory offered more relational and descriptive information, including perils covered, country and state/province coverage, trigger type, and risk modeler. These two sources were merged using composite fuzzy matching and subsequently validated to ensure the accurate consolidation of contract information. Descriptive statistics of the final dataset are presented in Tables \ref{tab1}--\ref{tab5}.

The dataset is characterized by high-cardinality categorical features. It includes 21 unique perils across 32 countries and 73 states/provinces, issued by 129 distinct cedents and underwritten by 32 unique underwriters. Representing these features in a standard tabular format, for instance via one-hot encoding, would create a prohibitively wide feature space of approximately 238 variables. Given the limited number of bond contracts, this high dimensionality introduces the `curse of dimensionality,' a problem that often plagues the performance of traditional machine learning models. Furthermore, alternative feature engineering techniques, such as label, frequency or target encoding, would collapse these features into simplified values, losing the nuanced relational information that is critical for pricing. In the next section, we introduce a graph representation of the CAT dataset, which is designed to thrive on the rich relational nature of this data.

\section{Graph representation of CAT data} \label{sec3}

The power of graph formalism lies not only in its focus on relationships between points, but also in its generalization. Unlike traditional data representations that emphasize individual data points, graphs prioritize the connections between them. This focus on relationships allows for a more nuanced understanding of complex systems.  

Beyond providing an elegant theoretical framework, graphs offer a robust mathematical foundation for analyzing, understanding, and learning from complex systems. Before delving into representational learning on graphs, we must first establish a more formal definition of a ``graph'' and understand its topology.  

A network, represented by a graph, consists of two main components-- nodes (vertices) and edges (links). Nodes represent the entities or units within the graph, while edges illustrate the connections between them. For instance, in a social network, individuals are represented as nodes, and their friendships are depicted by edges. More complex systems, such as a power grid (with nodes representing power plants and edges representing cables) or protein interactions (with nodes corresponding to proteins and edges representing binding interactions), can also be effectively modeled using graphs. It is important to note that two networks can have the same graph representation but differ in nature. To provide a more concrete definition, we denote a graph, $G$ by its sets of nodes and edges. Formally, a finite graph $G$ is a pair $(\mathcal{V}(G), \mathcal{E}(G))$, where $\mathcal{V}(G)$ is the countable vertex set (or node set) and $\mathcal{E}(G)$ is the edge set. To distinguish nodes, we label them with a subscript $i=1,2, \ldots, N$, where $N$ is the size of the graph (i.e., the number of nodes). Therefore, the node set $\mathcal{V}(G)$ can be represented as $\left\{v_1, v_2, \ldots, v_N\right\}$, with $v_i$ denoting the $i^{\text {th }}$ node in the graph. A comprehensive representation of a graph can be encapsulated in what is known as an adjacency matrix. The elements of the adjacency matrix show the direct connection between any pairs of nodes.  \\

\subsection{Multi-relational graph}\label{sec:sec3.1}

\begin{figure}[H]
    \centering
    \includegraphics[scale= 0.5]{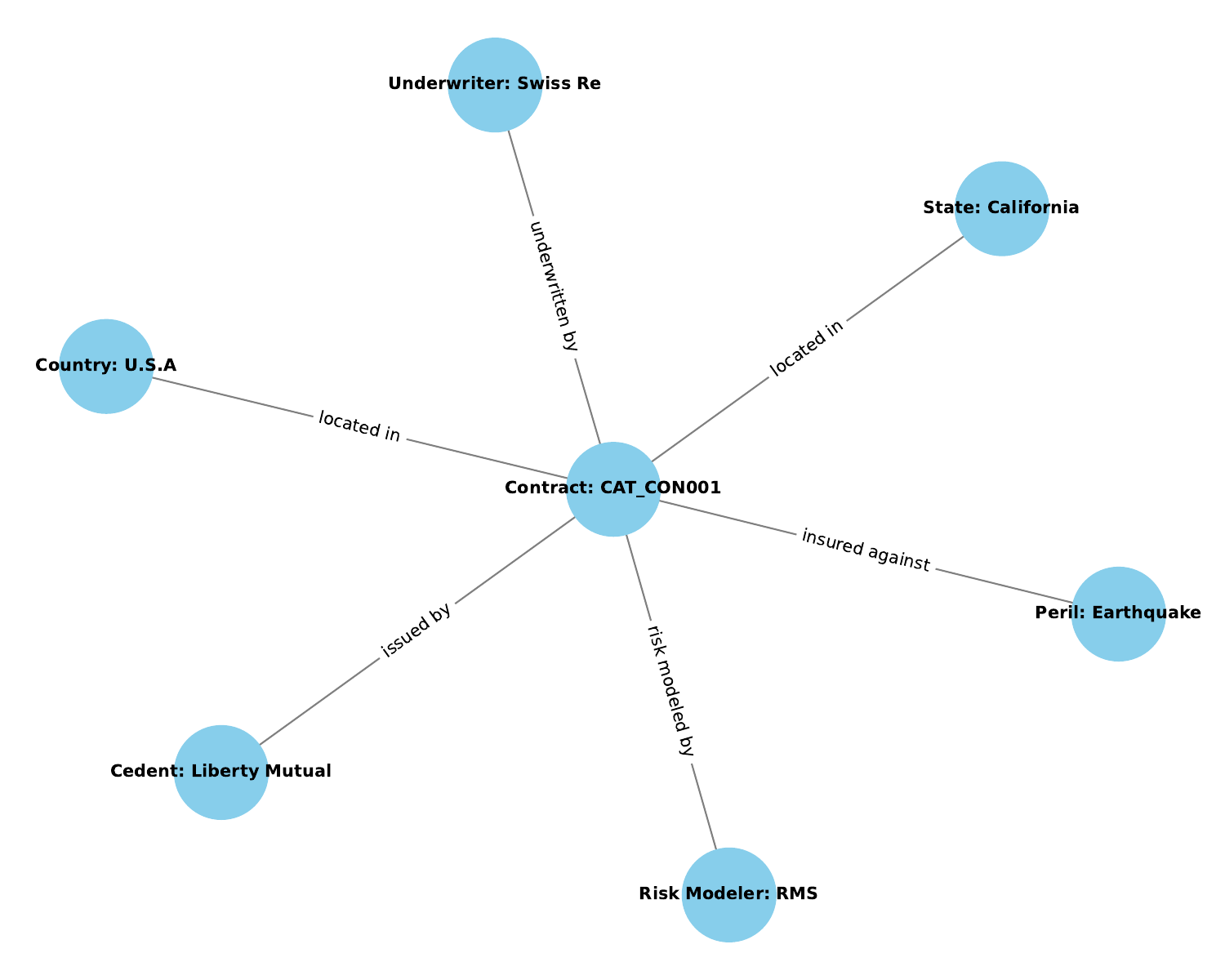}
    \caption{Relational graph representation of a typical CAT bond contract.}
    \label{relgraph}
\end{figure}

In this article, we represent the CAT bond data as a multi-relational graph (see Figure \ref{relgraph}). A multi-relational graph is a generalization of the standard graph, $G$ where the edges represent different types of relationships between nodes. This structure is particularly useful for modeling complex systems with multiple types of interactions, such as CAT bond contracts covering various perils across different regions. An example would be drug-drug interaction networks, where different edges correspond to various side effects between pairs of drugs. More formally, for each relation type $r \in \mathcal{R}$, where $\mathcal{R}$ is the set of all possible relations, we define an adjacency matrix\footnote{refer to definition in Appendix \ref{topo}} $\mathbf{A}_r$, such that each edge in the graph is described by a tuple $(u, r, v)$ belonging to the edge set $\mathcal{E}(G)$. The full structure of the graph can then be captured in a tensor $\mathcal{A}\in \mathbb{R}^{|\mathcal{V}(G)|\times |\mathcal{R}(G)|\times|\mathcal{E}(G)|}$, summarizing all relationships between the nodes. A multi-relational graph with different node types is called a heterogeneous multi-relational graph, in which case nodes can be divided into disjoint sets, such that $\mathcal{V}(G) = \mathcal{V}_1(G) \cup \mathcal{V}_2(G) \cup \dots \mathcal{V}_k(G)$, where $\mathcal{V}_i(G) \cap \mathcal{V}_j(G) = \emptyset$. 

As shown in Figure \ref{relgraph}, each node, $v \in \mathcal{V}$ represents an entity such as ``Country'', ``Underwriter'', ``State/Province'', ``Peril'', ``Risk Modeler'', and ``Cedent''. For each relation, $r \in \mathcal{R}$ represents a particular type of interaction between the CAT contract and the other entities. Each edge $(u, r, v)$ denotes the type of relationship, $r$ from node $u$ to node $v$. With such a structure in place, we can perform computations on the graph by leveraging its adjacency structure. For each relation $r \in \mathcal{R}$, we can define the adjacency matrix $\mathbf{A}_r \in \mathbb{R}^{|\mathcal{V}| \times|\mathcal{V}|}:$

\begin{equation}
\left(\mathbf{A}_r\right)_{u v}= \begin{cases}1, & \text { if }(u, r, v) \in \mathcal{E} \\ 0, & \text { otherwise }\end{cases}
\end{equation}

\subsection{Graph construction}\label{sec:sec3.2}

Following the multi-relational graph scheme outlined in Section \ref{sec:sec3.1}, we construct a CAT bond network to represent the complex relationships among entities involved in catastrophe bond issuance. This construction involves several key steps:

\textit{Node Creation:} We first identify the key entities in the dataset: isuers (cedents), geographic locations (country and state/province), peril names, underwriters, and risk modeling agencies, as depicted in Figure \ref{relgraph}. Each unique entity is represented as a node in the graph, with attributes assigned to indicate its type.

\textit{Edge Creation:} Next, we define the relationships between these entities based on their roles in the bond contracts. The primary bond identifier (e.g., CAT\_CON001) is connected to other entities, creating edges that signify direct relationships. Each edge is assigned a type that reflects the nature of the relationship, providing a structured understanding of how different entities interact within the market. Additionally, pairwise connections are created between entities within the same transaction, further enriching the relational structure of the graph.

\textit{Node Feature Enhancement:} To enhance node representations, we integrate additional contextual features from the dataset. Specifically, we assign all remaining CAT contract information, such as the S\&P rating of the bond, spread premium, expected loss, issue year, etc. (see Tables \ref{tab1}--\ref{tab5} for the full list), to the contractID node type. This inclusion ensures that nodes encapsulate important characteristics that may influence predictions and analyses.

\begin{figure}[H]
    \vspace{-0.30\textwidth}
    \hspace{-0.16\textwidth}
    \includegraphics[width=1.3\textwidth]{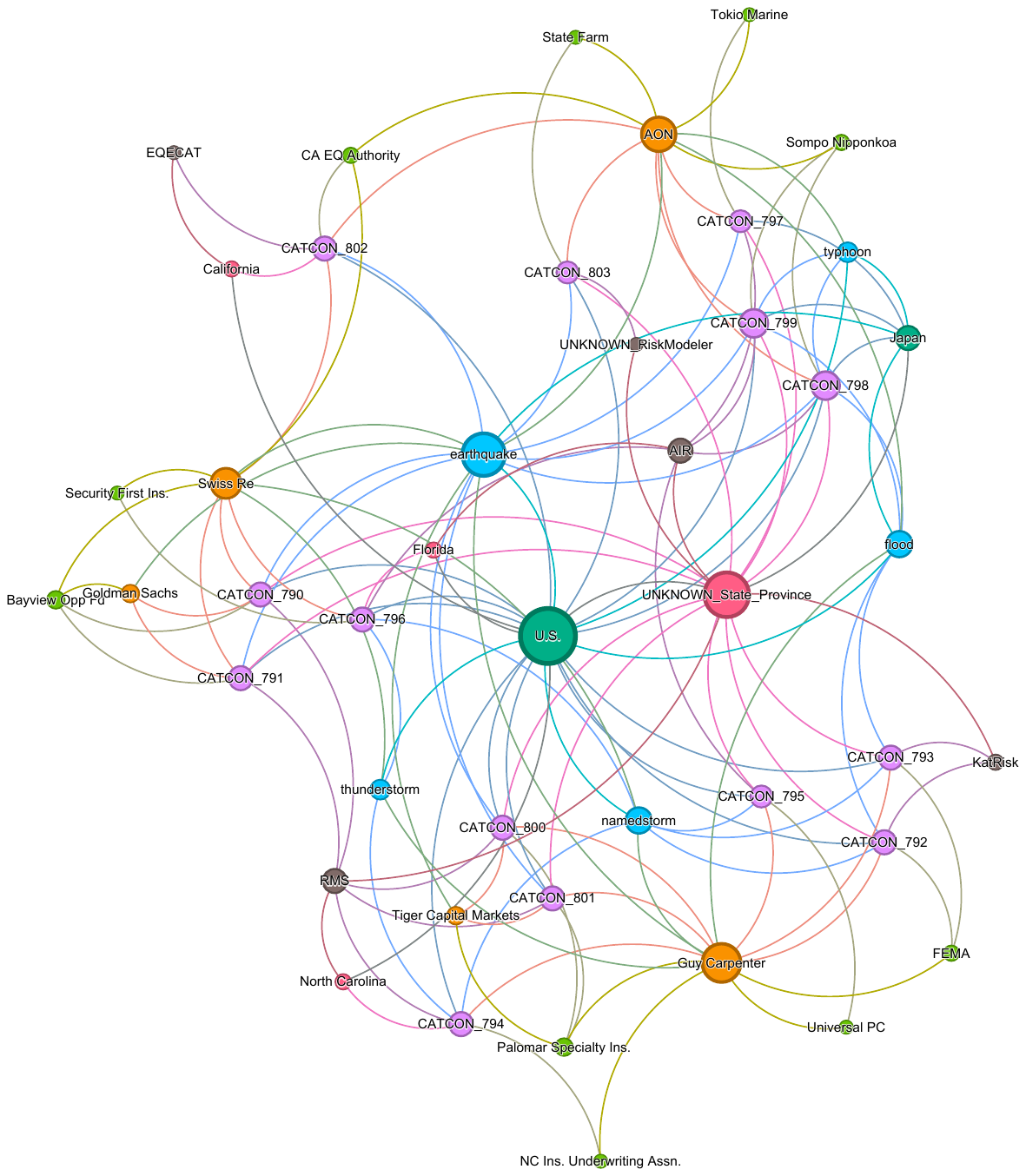}
    \vspace{-0.25\textwidth}
    \caption{CAT bond network visualization for bonds issued in 2021. Larger nodes indicate higher degrees. Different edge colors signify unique relationships between the entities.}
    \label{fig2}
\end{figure}

The resulting graph structure, containing 1,095 nodes and 49,464 edges (Figure \ref{fig3}), provides the foundation for our node-level analysis. Figure \ref{fig2} displays the CAT bond network for contracts issued in 2021. Although the dataset originates from 803 catastrophe bond contracts, the graph transformation creates an expansive relational environment that provides ample data for a deep learning model. This approach effectively addresses potential concerns regarding sample size; whereas a deep model would inevitably overfit on a tabular dataset of 803 rows, the R-GCN thrives on the rich topology of the graph. By shifting the learning task from isolated rows to interconnected entities, the model can extract nuanced features from the nearly 50,000 edges, ensuring high predictive performance without compromising model stability.

\subsection{CAT bond network properties}\label{sec:section3.3}

Analysis of the constructed CAT bond network offers key insight into the structure of the primary market.  We begin by examining the degree distribution of the nodes to determine if the relationships within the market are random \footnote{A random graph is a graph whose nodes  are connected in a random manner and as a result it has a random number of edges, refer to Appendix \ref{topo} for more technical detail.}. The degree of a node, defined as the number of edges connected to it, indicates the number of direct relationships a node has with other nodes in the graph. 

\begin{figure}[!hb]
    \centering
    \includegraphics[scale= 0.7]{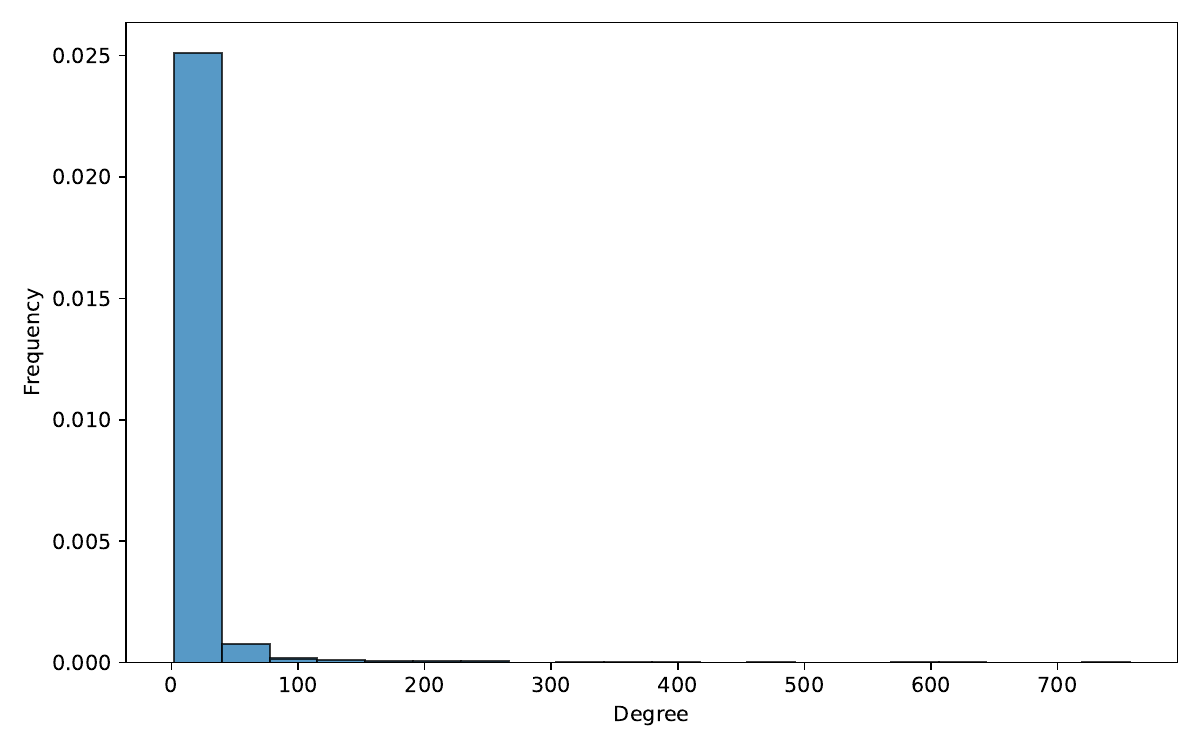}
    \caption{Histogram plot of degree frequency}
    \label{fig4}
\end{figure}

Figure \ref{fig4} illustrates the highly skewed degree distribution of the CAT bond network, with most nodes exhibiting between 0 and 50 connections. However, the presence of several highly connected nodes suggests that the network may follow a power-law distribution, a characteristic often observed in real-world networks.  We further investigate this possibility in the subsequent analysis.

A commonly used degree distribution model is the fat-tailed power-law distribution. However, empirical analyses of numerous real-world networks reveal that a strict adherence to a pure power-law distribution may not accurately capture the characteristics of real networks. Consequently, an adjusted power-law distribution has been proposed, characterized by two additional parameters known as low-degree saturation ($k_{sat}$) and high-degree cutoff ($k_{cut}$). These parameters serve to reconcile the observed lower frequency of low-degree and high-degree nodes with what a pure power-law distribution would predict. Thus, the following model is fitted to the degree sequence: 
\begin{equation}
    p_k (\gamma, k_{sat}, k_{cut})= \frac{(k + k_{sat})^{-\gamma}}{  \sum_{k^\prime}(k^\prime + k_{sat})e^{-\frac{k^\prime}{k_{cut}}}  }\exp\left(-\frac{k}{k_{cut}}\right)
\end{equation}
The determination of the optimal parameters $k_{sat}$ and $k_{cut}$ is conducted by iteratively scanning values within the range of $k_{min}$ to $k_{max}$, minimizing the Kolmogorov-Smirnov test statistic\footnote{The iteration process is initialized with estimating $\gamma$ using the following relation for a given value $K^* \in [k_{sat}, k_{cut}]$, a point beyond which the data behave exactly the power-law distribution: 
\begin{equation}
    \gamma = 1+N\left[ \sum_{i=1}^{N}  \log \frac{k_i}{K^* - \frac{1}{2}}  \right]^{-1}
\end{equation}
}. Next, the degree exponent $\gamma$ is found by maximizing the log-likelihood function given by 
\begin{equation}
    \log L(\gamma; k_{sat}, k_{cut}) = \sum_{i=1}^{N}\, \log\, p_{k_i}(\gamma, k_{sat}, k_{cut})
\end{equation}
Subsequently, a bootstrapping technique is employed to ascertain the $p$-value associated with the goodness-of-fit test.  

\begin{table}[!ht]
    \centering
    \begin{tabular}{c c c c c}
    \toprule
      Distribution &  $\gamma$ & $k_{sat}$ & $k_{cut}$ & $p$-value \\[0.1cm]\hline
        Adjusted Power-Law  &  2.033 & 44  & 607 & 0.94 \\[0.1cm]
        \bottomrule
    \end{tabular}
    \caption{Estimates of the power-law distribution parameters.}
    \label{tab6}
\end{table}

Table \ref{tab6} demonstrates that a power-law distribution effectively models the observed node degrees, as evidenced by the bootstrapped $p$-value of the goodness-of-fit test. The degree exponent ($\gamma$), falling within the range of 2 to 3, further suggests that the CAT bond network exhibits characteristics of a scale-free network. This observation has important implications for understanding the issuance of CAT bonds in the primary market.

A scale-free financial network is characterized by a small number of highly connected entities with a large number of connections, while the majority of nodes have few connections. These highly connected nodes, often referred to as ``hubs," play a crucial role in the network, but their dominance also creates vulnerabilities.

The concentration of connectivity in a scale-free network can lead to concentrated risk. Disturbances affecting these key nodes can propagate rapidly, potentially triggering cascading failures and systemic crises. While these hubs facilitate efficient information and liquidity transmission during normal market conditions, they can exacerbate fragility during times of stress.

\begin{figure}[!hb]
    \centering
    \includegraphics[scale= 0.6]{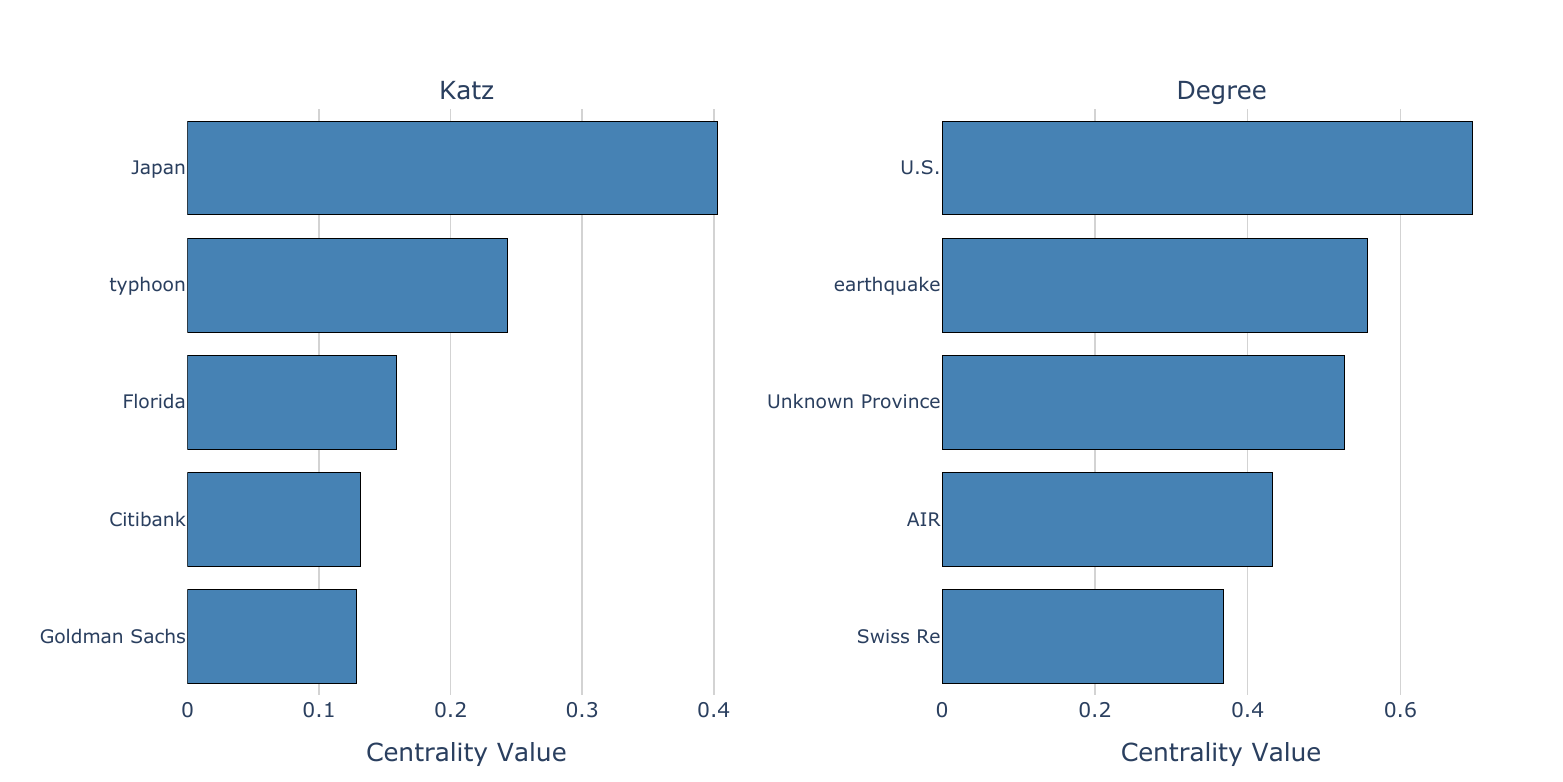}
    \caption{Centrality measures by node (top 5 per measure)}
    \label{fig5}
\end{figure}

In the context of CAT bonds, the high-degree nodes represent entities involved in the majority of CAT bond contracts (in different capacities) issued in the primary market. Figure \ref{fig5} highlights the most important nodes in terms of degree centrality: ``U.S.," ``earthquake," ``AIR," and ``Swiss Re." This underscores the concentration of CAT bond insurance coverage in the U.S., with earthquakes as a dominant peril (primarily due to the significant exposure of outstanding bonds to U.S. hurricanes and earthquakes), and AIR and Swiss Re serving as the primary risk model provider and cedent, respectively. The dominance of Swiss Re (stemming from its dual role as an issuer and underwriter) as a central hub in the CAT bond network may raise concerns about systemic risk. If Swiss Re were to experience financial distress or a significant reduction in its underwriting capacity, it could disrupt the entire CAT bond market, potentially limiting the availability of catastrophe insurance coverage and hindering the transfer of risk from insurers to investors. This concentration of activity within a single entity highlights a potential vulnerability in the CAT bond market, underscoring the need for diversification and the careful monitoring of key players.
 
However, the Katz index \footnote{The Katz index is defined as the weighted sum of all possible paths between two nodes. For further details, refer to appendix \ref{topo}. Unlike degree centrality, which considers only the local neighborhood, the Katz index captures connectivity across the entire network.} tells a different story. The initial analysis of the data reveals that the high Katz index values observed for the nodes in Figure~\ref{fig5} stem from their direct or indirect connections to key hubs (high-degree nodes) associated with critical factors such as the U.S., earthquakes, and hurricanes.

For instance, Japan is frequently grouped with the U.S. and Europe, both of which have high-risk exposures and encompass the majority of the CAT market outstanding. This bundling is advantageous because it allows for geographical risk diversification. California, as the most earthquake-prone region in the U.S., and Tokyo, as the most earthquake-prone region in Asia, can leverage this diversification to mitigate localized risks. However, the concentration of risk in the U.S., coupled with the severity of events that occur there, increases the likelihood that a contract triggered by a U.S. event could propagate risk to other covered areas.

Similarly, Citibank, as an underwriter node type, is often linked with other influential underwriters such as Goldman Sachs, Swiss Re, and AON, all of which play significant roles in underwriting catastrophe bonds. Notably, Citibank frequently collaborates with these key entities, particularly in contracts covering the U.S., a region with high exposure to catastrophic events. This consistent participation and association with high-risk regions explain Citibank’s elevated Katz index value.

These observations provide valuable insights into structuring effective catastrophe bond deals. To mitigate risk exposure, contracts may focus on maximizing geographical diversification while leveraging the reputation and capabilities of well-established entities, including risk modelers, cedents, and underwriters. However, a contract that includes a representative level from the node type with the highest Katz index, if not carefully structured, may inadvertently increase risk exposure, making it a less desirable option.

To study the dynamics of the CAT bond network over time, we calculate the fitness of each node, which measures the tendency of new nodes to connect to an existing node in the network \footnote{In network science, the tendency for new nodes to in a network to connect themselves to existing nodes that already have high degree (i.e., nodes that are well connected) is called preferential attachment.}. Higher fitness indicates a greater likelihood that a node’s degree will increase as the network grows. 

The dynamics of the CAT network can be observed by analyzing sub-graphs that are expanded based on the ``issue year'', starting from 1999 through 2021. For each year $(i=1999,…,2021)$, we consider a sub-graph corresponding to the network of CAT contracts with an ``issue year'' of 
$i$. By continuing this process for each subsequent year, we can track the network’s evolution over the entire period.

\begin{figure}[!hb]
    \centering
    \includegraphics[scale= 0.57]{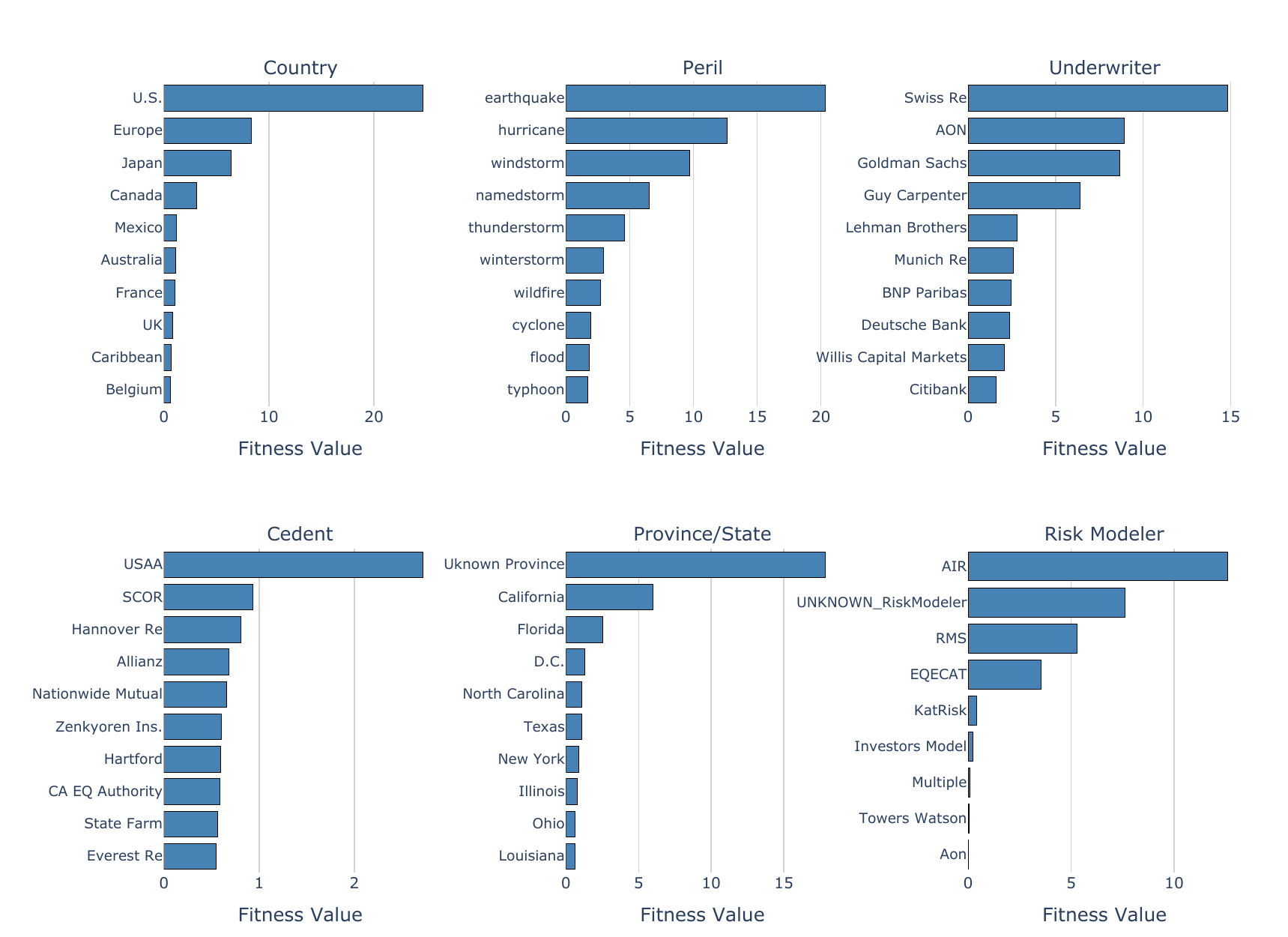}
    \caption{Fitness of different node types.}
    \label{fig21}
\end{figure}

Figure \ref{fig21} depicts the top 10 nodes with the highest fitness values. The plots confirm that there is a tendency for new CAT bond issuance to be within entities which already have most volumes. As an example, the U.S. continues to be the main destination of CAT issuance. However, aside from Japan and the U.S., where most CAT bond placements are concentrated, France and Belgium in Europe show potential to attract more investor attention in the coming years. Storm-related events, in particular, are becoming increasingly appealing as the CAT market grows, and we can expect to see more CAT bonds covering these types of disasters in the future.

Another implication of the scale-free property is that the network structure is robust against random node removal but sensitive to selective removal.\footnote{This area of research in network science is referred to as robustness analysis, where network resilience, often measured by the relative size of the giant component, is tested by removing nodes either randomly or selectively. As the name suggests, in a random removal, nodes are removed in a random manner. On the other hand, selective removal can be based on various criteria, such as degree centrality, where nodes with the highest degree are removed first, followed by the next highest, and so on.} In the context of CAT bonds, node removal, whether random or selective, corresponds to situations where entities such as cedents default, or specific characteristics are removed from the contract. Under the random removal scheme, the critical threshold (a fraction of the nodes that must be removed for the giant component to break down) is estimated as $f_c = 1-(\frac{1}{\langle k^2 \rangle /\langle k \rangle -1}) = 0.9933$. This high level of robustness suggests that the network is minimally affected by random node removal. In other words, missing information has a limited impact on the performance of a model built on a GNN, making it resilient to incomplete data. However, we note that the network is vulnerable to the selective removal of hubs. This selective removal, which has significant financial implications, can be seen as analogous to a cedent with high risk exposure going bankrupt. \\

The Pearson coefficient correlation of $r = -0.397$ for the CAT network indicates a negative correlation between the degrees of connected nodes. In essence, the CAT network exhibits a hierarchical, ``hub-and-spoke'' topology, where a few hub nodes of high degree connect to numerous low-degree peripheral nodes. 

\begin{figure}[H]
    \centering
    \includegraphics[scale= 0.7]{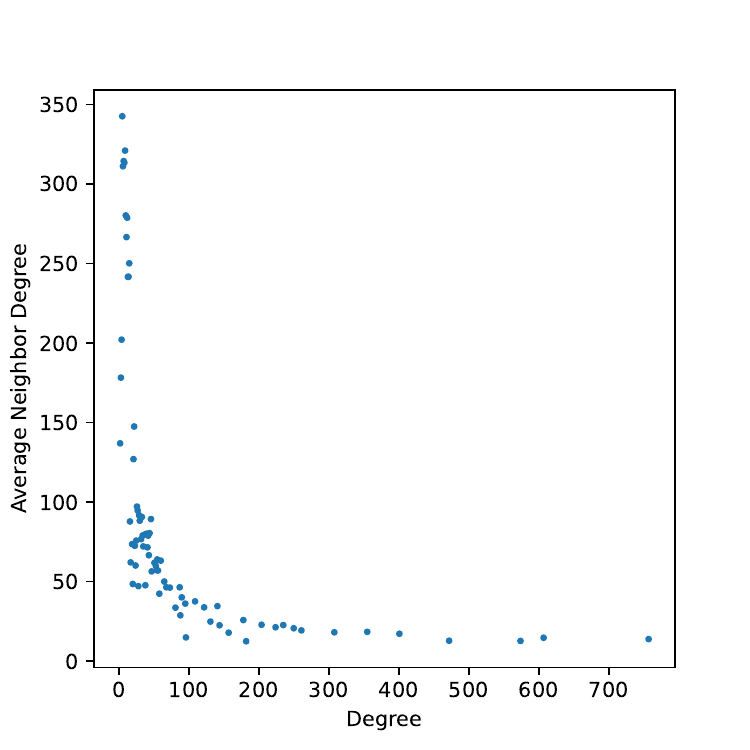}
    \caption{The average neighbor degree across all $k$ degree nodes as a function of degree $k$}
    \label{fig8}
\end{figure}

Additionally, Figure \ref{fig8} displays the average neighbor degree across all $k$ degree nodes as a function of degree $k$, revealing a decreasing trend for which the Pearson coefficient correlation equal to $-0.466$\footnote{We make a note that the statistical significance of results is checked using a two-sided t-student test at a confidence level of $95$ percent.}. These findings show that the CAT network tends to be a disassortative network\footnote{A network where low-degree nodes tend to connect high-degree nodes, see appendix \ref{topo} for more information. } due to the negative correlation coefficient. The implication is that high-degree hubs have neighbors with substantially lower degree on average, and low-degree nodes tend to link to higher-degree neighbors. This disassortative phenomenon further corroborates our finding on the scale-free nature of CAT network previously discussed. 

We finalize the network analysis by providing a summary of CAT network's properties, outlined in Table \ref{tab7}.

\begin{table}[H]
  \centering
  \caption{CAT Network Properties}
  \label{tab7}
  \begin{tabular}{@{}ll@{}}
    \toprule
    Property                          & Value     \\ \midrule
    Number of Edges ($L$)             & 49464      \\
    Number of Nodes ($N$)             & 1095      \\
    Average Degree ($\langle G \rangle$) & 15.52   \\
    The second moment ($\langle G^2 \rangle $) & 2346.50 \\
    Minimum Degree ($k_{\min}$)       & 2         \\
    Maximum Degree ($k_{\max}$)       & 757       \\
    Diameter ($\text{diam}\langle G\rangle$)       & 5         \\
    Average Path ($\langle \text{dist}_G \rangle$) & 2.18 \\
    Average Clustering Coefficient ($\langle C \rangle$) & 0.38 \\
    Global Clustering Coefficient ($C_{\Delta}$) & 0.021 \\
    Network type & Undirected \\
    Connected Network & Yes \\
    Assortative Network & No \\
    Power-law degree distribution & Yes\\
    Scale-free Network & Yes \\
    Small-world Network & Yes\\ \bottomrule
  \end{tabular}
\end{table}

\section{Geometric Deep Learning for node prediction task}\label{sec4}

Geometric Deep Learning (GDL) extends the capabilities of deep learning to non-Euclidean domains like graphs and manifolds \citep{Bronstein2017}. While traditional deep learning models such as Convolutional Neural Networks (CNNs) and Recurrent Neural Networks (RNNs) excel at processing grid-like structures (e.g., images, sequences), many real-world datasets, particularly in finance and social networks, are inherently graph-structured. GDL provides the necessary framework for neural networks to effectively learn from this type of data.

\subsection{Graph neural networks}
\begin{figure}[H]
    \centering
    \includegraphics[scale= 0.7]{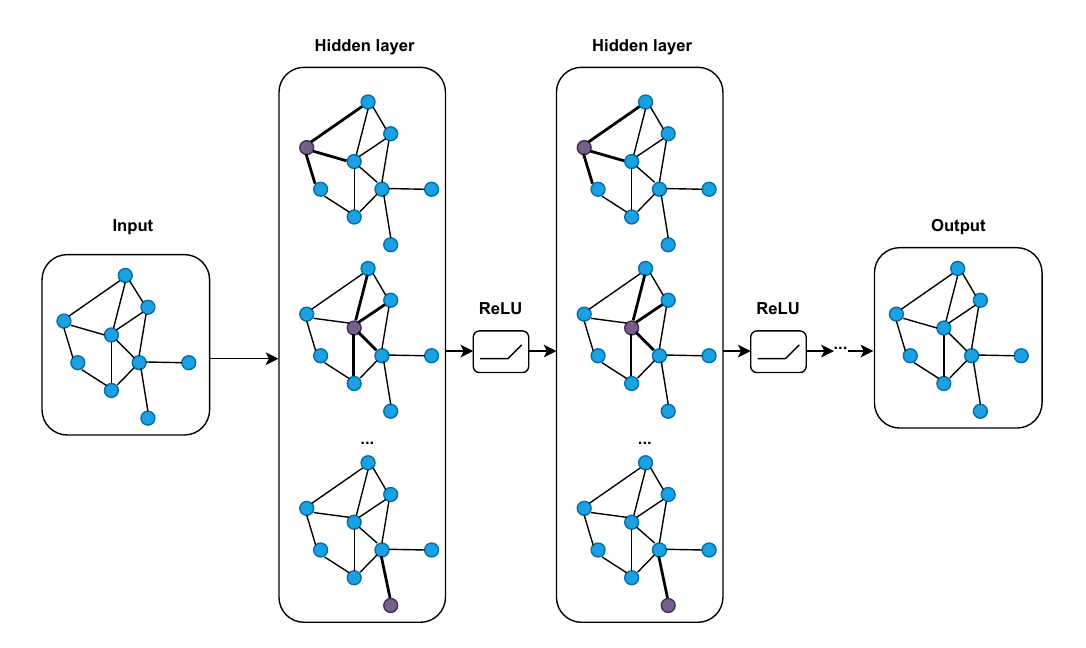}
    \caption{High-level GNN architecture showing how node features and graph topology are integrated for learning}
    \label{fig23}
\end{figure}

Graph Neural Networks (GNNs), a key component of GDL, are specifically designed to operate directly on graph structures \citep{Kipf2017}. These networks leverage both the topology of the graph and the features of its nodes to learn meaningful representations (see Figure~\ref{fig23}).

\begin{figure}[H]
    \centering
    \captionsetup{font=footnotesize}
    \includegraphics[scale= 0.8]{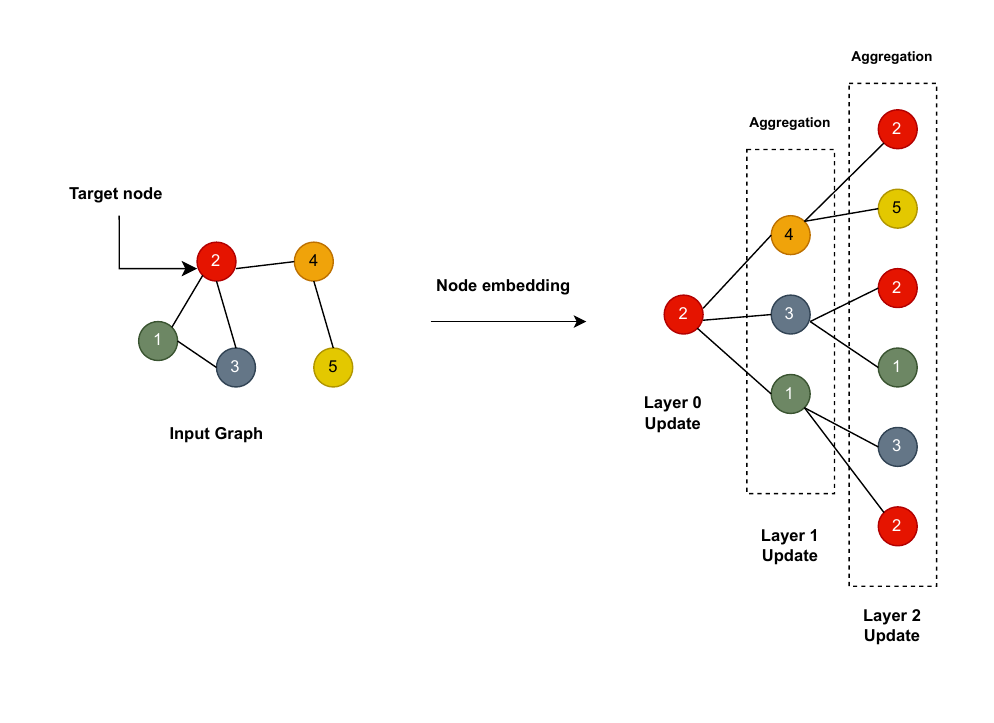}
    \vspace{-0.05\textwidth}
    \caption{Schematic of the message-passing process, where each node, representing an entity such as a CAT bond, peril type, or market participant, updates its representation by aggregating information from connected entities over multiple iterations. This approach enables the model to capture both local structural dependencies and cross-entity interactions.}
    \label{fig50}
\end{figure}

The core mechanism behind GNNs is the iterative updating of each node's representation by aggregating information from its neighbors, a process known as \textit{message passing}. Figure \ref{fig50} shows that schematic view of the aggregation process.

The message passing framework can be formalized as:
\begin{equation}\label{eq:4.1}
\mathbf{h}_u^{(k+1)}=\operatorname{UPDATE}^{(k)}\left(\mathbf{h}_u^{(k)}, \operatorname{AGGREGATE}^{(k)}\left(\left\{\mathbf{h}_v^{(k)}: v \in \mathcal{N}(u)\right\}\right)\right)
\end{equation}
where $\mathbf{h}_u^{(k)}$ represents the hidden state (embedding) of node $u$ at layer $k, \mathcal{N}(u)$ denotes the neighborhood of node $u, \operatorname{AGGREGATE~}^{(k)}$ is the function that aggregates messages from neighbors, and $\mathrm{UPDATE}^{(k)}$ is the function that updates the node's embedding. Initialization is performed as $\mathbf{h}_u^{(0)}=\mathbf{x}_u$ where $\mathbf{x}_u$ is the initial feature vector of node $u$. This framework ensures that the learned representations are permutation invariant and effectively capture local graph structures.

\subsubsection{Relational graph convolutional networks (R-GCNs)}

In many applications, graphs exhibit multiple types of relationships between nodes, resulting in multi-relational graphs. For instance, in the context of CAT bonds, nodes could represent contracts, perils, regions, and issuers, while edges capture diverse relations such as ``covers,'' ``issued by,'' or ``located in'' as already described in Section \ref{sec:sec3.1}. R-GCNs extend the capabilities of GNNs to effectively handle such multi-relational data \citep{Schlichtkrull2018}. They achieve this by introducing relation-specific transformations in the \textit{message passing} process, allowing the model to discern and capture the unique semantics of different edge types.

For R-GCNs, the update function used for node embeddings in Equation \ref{eq:4.1} is modified as follows:
\begin{equation}
\mathbf{h}_u^{(k+1)}=\sigma\left(\sum_{r \in \mathcal{R}} \sum_{v \in \mathcal{N}_r(u)} \frac{1}{c_{u v r}} \mathbf{W}_r^{(k)} \mathbf{h}_v^{(k)}+\mathbf{W}_0^{(k)} \mathbf{h}_u^{(k)}\right)
\end{equation}

where $\mathcal{R}$ represents the set of all relation types, $\mathcal{N}_r(u)$ denotes the set of neighboring nodes connected to $u$ via relation $r, \mathbf{W}_r^{(k)}$ is the weight matrix for relation $r$ at layer $k, \mathbf{W}_0^{(k)}$ is the weight matrix for the self-loop (to incorporate $u$'s own features), $c_{u v r}$ is a normalization constant, and $\sigma$ is an activation function, such as ReLU. For each relation $r$, messages from neighbors are aggregated:

\begin{equation}\label{eq:4.3}
\mathbf{m}_{\mathcal{N}_r(u)}^{(k)}=\sum_{v \in \mathcal{N}_r(u)} \frac{1}{c_{u v r}} \mathbf{W}_r^{(k)} \mathbf{h}_v^{(k)}
\end{equation}
The normalization constant $c_{u v r}$ prevent features from nodes with high degrees from dominating the embeddings and also aides model convergence. A common choice is $c_{u v r}=\left|\mathcal{N}_r(u)\right|$, the number of neighbors connected via relation $r$. Node embeddings are updated by combining messages from all relations and applying non-linearity:
\begin{equation}\label{eq:4.4}
\mathbf{h}_u^{(k+1)}=\sigma\left(\sum_{r \in \mathcal{R}} \mathbf{m}_{\mathcal{N}_r(u)}^{(k)}+\mathbf{W}_0^{(k)} \mathbf{h}_u^{(k)}\right)
\end{equation}
To manage the increased number of parameters due to multiple relations, R-GCNs employ a parameter sharing (regularization) technique known as basis decomposition \footnote{Excessive parameters in a model can lead to overfitting, increased computational burden, and slower learning due to the high dimensionality. The basis decomposition approach uses shared parameters $\mathbf{B}_b^{(k)}$, to capture common patterns across all relations, while also incorporating relation-specific weights, $a_{r b}^{(k)}$, to allow for variations specific to each relation, $r$.}:
\begin{equation}\label{eq:4.5}
\mathbf{W}_r^{(k)}=\sum_{b=1}^B a_{r b}^{(k)} \mathbf{B}_b^{(k)}
\end{equation} with typically $\mathbf{B} \ll|\mathcal{R}|$. $\mathbf{B}_b^{(k)}$ represents the basis matrices shared across relations and $a_{r b}^{(k)}$ are the coefficients specific to relation $r$. This decomposition effectively reduces the number of parameters from $\mathcal{O}(|\mathcal{R}| \times d \times d)$ to $\mathcal{O}(B \times d \times d)$, where $B$ is the number of bases and $d$ is the embedding dimension.

By incorporating the basis decomposition (Equation \ref{eq:4.5}) into the message passing function in Equation \ref{eq:4.3}, the final update to the message passing function becomes: 
\begin{equation}\label{eq:4.6}
\mathbf{h}_u^{(k+1)}=\sigma\left(\sum_{r \in \mathcal{R}} \sum_{v \in \mathcal{N}_r(u)} \frac{1}{c_{u v r}}\left(\sum_{b=1}^B a_{r b}^{(k)} \mathbf{B}_b^{(k)}\right) \mathbf{h}_v^{(k)}+\mathbf{W}_0^{(k)} \mathbf{h}_u^{(k)}\right)
\end{equation}

During the forward pass of an R-GCN, messages(information) are passed along the edges that connects different types of nodes. This means that the features of the neighboring nodes are aggregated and used to update the features of the target node (see Figure~\ref{fig50}). The R-GCN learns distinct weights for each relation type, allowing it to handle the heterogeneous nature of the graph. 

\subsubsection{R-GCN application to CAT bond risk premium prediction}

Since our goal is to predict the risk premiums of CAT bonds by capturing the complex relationships among the various entities involved, we constructed a multi-relational graph as detailed in Section \ref{sec:sec3.2}. We initialize node embeddings with $\mathbf{h}_u^{(0)}=\mathbf{x}_u$, where $\mathbf{X}_u$ represents the node features (e.g., bond attributes). Multiple R-GCN layers are then applied to propagate and transform information across the graph (see Figure~\ref{fig20}).
For each contract node $u$, the final embedding is computed as $\mathbf{z}_u=\mathbf{h}_u^{(K)}$. This embedding is then used to predict the risk premium, $\hat{y}_u$ using a regression function:
\begin{equation}
\begin{aligned}
&\hat{y}_u=\mathbf{w}^{\top} \mathbf{z}_u+b\
\end{aligned}
\end{equation}
where $\mathbf{w}$ and $b$ are the parameters of the regression function, learned by the model during training. Finally, the mean squared error (MSE) between the predicted and true risk premiums is calculated as:
\begin{equation}
\mathcal{L}=\frac{1}{|\mathcal{D}|} \sum_{u \in \mathcal{D}}\left(y_u-\hat{y}_u\right)^2
\end{equation}
where $\mathcal{D}$ represents the set of contract nodes in the training data, and $y_u$ is the true risk premium of contract $u$.

\subsection{R-GCN Implementation}
This section outlines the data preprocessing and feature transformation techniques employed to prepare the CAT bond data for use with R-GCNs. 

\subsubsection{Data transformation}

\textit{Temporal feature extraction and transformation:}
The reinsurance market exhibits cyclical behavior, characterized by alternating hard and soft market periods \citep{WeissChung2004}. Hard markets are characterized by diminished reinsurance capacity, leading to higher premiums and reduced CAT bond issuance volumes. Conversely, soft markets exhibit abundant capacity and increased issuance. Traditionally, researchers have used the Rate-on-Line (RoL) index to capture these market cycles \citep{Cummins2009}. 
In this implementation, we capture cyclical effects and market regimes directly from the issue dates of CAT bonds. The premise is that the issue date inherently reflects market conditions, as it influences and is influenced by the prevailing market regime. By decomposing the issue date into meaningful components, we extract temporal patterns that serve as significant predictors of risk premiums.

We define \textit{epoch} as the earliest issue year in the dataset and calculate the number of years since the epoch for each bond issuance.  The cyclical nature of months is encoded using \textit{sine} and \textit{cosine} transformations, ensuring that December and January are numerically adjacent. This transformation allows the model to learn patterns associated with market regimes over time.

By transforming date features to capture cyclical market behaviors, we integrate temporal dynamics directly into the model without relying on external datasets like RoL. This ensures that the model bases its predictions solely on information embedded within the bond issuance data, aligning with the efficient market hypothesis, which suggests that all available information, including market regime indicators, is reflected in bond prices at the time of issuance \citep{Fama1970}.

\textit{Feature standardization and transformation:} To prepare the data for the R-GCN model, we encode categorical variables, particularly those with multiple values per observation, and standardize numerical features to improve model convergence. Categorical variables with two sets of unique values are converted into binary features using one-hot encoding. For columns containing lists of categories (e.g., Trigger Types, S\&P Rating), binary features are created for each unique category across all entries.
 
\subsubsection{Quantifying structural influence via topological diagnostics}\label{sec:sec4.2.2}

A central premise of CATNet is that by representing the catastrophe bond market as a relational graph, we eliminate the need for manual feature engineering and external macroeconomic data. In this paradigm, the model's predictive power stems from the inherent topology of the network itself. Unlike tabular models that treat contracts as independent observations, the R-GCN architecture captures the ``connectedness'' of the market participants through its message-passing mechanism. 

\begin{figure}[H]
    \centering
    \includegraphics[scale= 0.57]{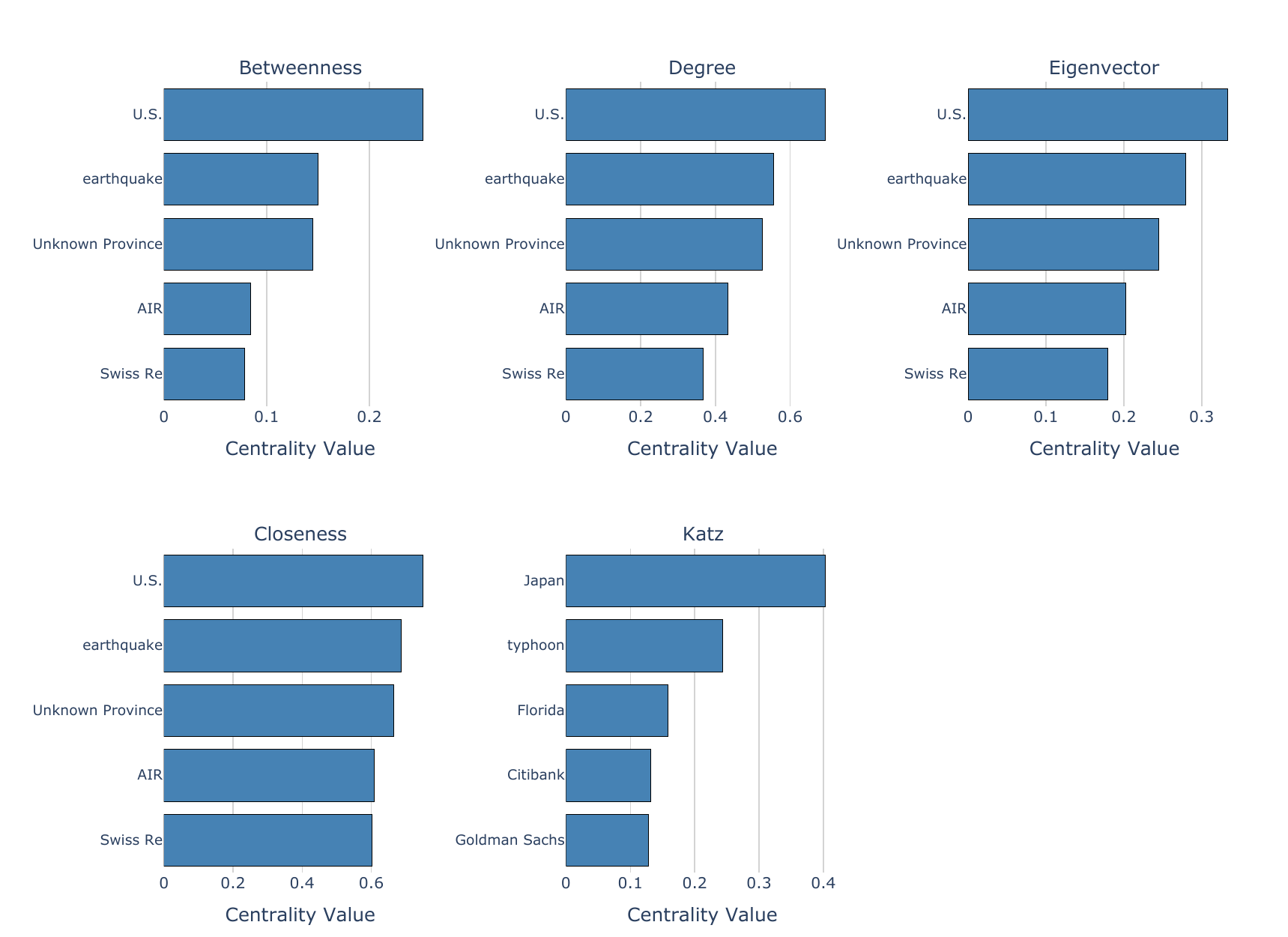}
    \caption{Centrality measures by node (top 5 per measure)}
    \label{fig20}
\end{figure}

To illustrate the quantitative significance of this structural information, we use six centrality measures as topological diagnostics (see Figure \ref{fig20}). It is important to distinguish these from traditional feature engineering; these measures are not external inputs but are mathematical byproducts of the graph's existing node-edge relationships. By explicitly including these metrics in our node importance analysis (see  Section \ref{Section:topo}, Figure \ref{fig9}), we can ``unmask'' the latent relational patterns that the R-GCN exploits.

For diagnostic purposes, we initially employed these features in a unidirectional graph setting, where they provided a significant performance boost by compensating for limited relational flow. In our more advanced bidirectional implementation, we observed that these explicit features became redundant, as the model's enhanced message-passing layers learned the equivalent structural information autonomously. We include them in our reported feature importance rankings to provide transparency into the ``black box'' of the GNN, demonstrating that market influence—proxied by metrics like Betweenness and Closeness Centrality—is a statistically significant determinant of CAT bond spreads. For a more formal definition of these centrality measures, refer to Appendix \ref{topo}

\subsubsection{Training: inductive versus inductive learning paradigms on graphs} \label{sec:sec4.2.3}

In graph representational learning, there are two canonical training paradigms: transductive and inductive. In a transductive setting, a model learns from the entire graph, including the features and connections of all nodes, and is evaluated on nodes it has already ``seen" during training. This approach specializes the model to a single, fixed graph structure. In contrast, an inductive setting trains a model to generalize to entirely new nodes or graphs. Instead of learning a unique embedding for each node, an inductive model learns an aggregation function that can generate embeddings for new entities on the fly. 

By design, the R-GCN architecture is transductive \citep{A8}. It learns a unique embedding for each node and a specific weight matrix for each relation type. Consequently, adding a new node to the graph is problematic, as the dimensions of the node embedding matrix ([N\_nodes, embedding\_dim]) would no longer align, typically requiring a full retraining of the model. 

This transductive nature, however, is well-suited to the catastrophe bond domain. As established in Section \ref{sec2}, the market is characterized by a finite set of recurring entities (e.g., perils, cedents, underwriters). Therefore, a new bond contract will almost certainly connect to existing nodes in the graph, allowing the model to leverage already-learned information to make predictions about the new transaction. 

We employ a hybrid testing strategy for out-of-sample (OOS) and out-of-time (OOT) evaluations based on a transductive setup with masked test features. To ensure the model learns from the graph’s topology rather than relying on feature leakage, the feature vectors of test nodes are zeroed out during training while their structural positions are preserved. At evaluation, the true features are reintroduced, and the model predicts spreads using frozen weights. This methodology allows us to evaluate the model's generalization capabilities without altering the underlying node embedding matrix. By using this approach, we leverage the relational strengths of the R-GCN architecture while transparently addressing its limitations in fully inductive environments where entirely new entities might appear.

Refer to Section~\ref{hyperparameters} for the hyper parameter tuning process of the R-GCN model using Optuna.

\section{Results \& Discussion} \label{sec5}
This section evaluates the performance of the R-GCN model in predicting catastrophe bond spreads against two robust tabular benchmarks: Random Forest (RF) \citep{A3} and XGBoost \citep{Chen2024}. Our central hypothesis is that the R-GCN's graph-based structure can more effectively capture the complex, relational information inherent in bond contracts than a tabular model. To test this and identify key performance drivers, we conduct ablation studies by including or excluding network-derived features. We also discuss in detail the features that drive the model's prediction.

\subsection{Out-of-sample performance evaluation}
We generated 10 random subsets of the data, each with an 80\% training and 20\% testing split. The models were trained and tested on each subset, and the average performance is reported in Table \ref{tab:oos_r2}.  The R-GCN model consistently outperformed established tabular benchmarks, achieving the highest average $R^2$ of 79.05\%, compared to 77.63\% for XGBoost and 73.09\% for Random Forest. Notably, the R-GCN recorded the lowest average error (MSE: 0.2109), representing a significant 22.0\% reduction in error relative to the Random Forest baseline and a 7.7\% reduction compared to XGBoost.

The R-GCN’s superior performance is attributed to its ability to leverage the market's underlying relational structure—specifically the connections between bonds sharing common market participants or peril types. This structural information serves as an implicit regularizer, allowing the model to generalize more effectively across the heterogeneous CAT bond universe than models limited to independent feature mappings. While tree-based models like XGBoost may remain competitive in instances where relational connectivity is sparse, the R-GCN provides a more robust framework for capturing the complex interdependencies of the primary market.

\begin{table}[htbp]
    \centering
    \caption{OOS model performance across folds}
    \begin{tabular}{lcccccc}
        \toprule
        \multirow{2}{*}{\textbf{Fold}} & \multicolumn{2}{c}{\textbf{Random Forest}} & \multicolumn{2}{c}{\textbf{XGBoost}} & \multicolumn{2}{c}{\textbf{R-GCN}} \\
        \cmidrule(lr){2-3} \cmidrule(lr){4-5} \cmidrule(lr){6-7}
        & \textbf{MSE} & \textbf{$\mathbf{R^2}$ (\%)} & \textbf{MSE} & \textbf{$\mathbf{R^2}$ (\%)} & \textbf{MSE} & \textbf{$\mathbf{R^2}$ (\%)} \\
        \midrule
        1 & 0.4345 & 61.21 & 0.3588 & 67.97 & 0.2754 & 75.42 \\
        2 & 0.1680 & 79.74 & 0.1573 & 81.02 & 0.1088 & 86.88 \\
        3 & 0.3051 & 73.92 & 0.2506 & 78.58 & 0.2849 & 75.64 \\
        4 & 0.2218 & 73.69 & 0.2024 & 75.99 & 0.2472 & 70.67 \\
        5 & 0.2199 & 64.72 & 0.1422 & 77.19 & 0.1231 & 80.26 \\
        6 & 0.3610 & 74.09 & 0.2737 & 80.36 & 0.2419 & 82.64 \\
        7 & 0.2262 & 80.72 & 0.2605 & 77.81 & 0.1958 & 83.31 \\
        8 & 0.3275 & 69.91 & 0.3013 & 72.32 & 0.2743 & 74.80 \\
        9 & 0.1842 & 75.79 & 0.1101 & 85.52 & 0.1671 & 78.03 \\
        10 & 0.2549 & 77.07 & 0.2275 & 79.54 & 0.1907 & 82.85 \\
        \midrule
        \textbf{Average} & \textbf{0.2703} & \textbf{73.09} & \textbf{0.2284} & \textbf{77.63} & \textbf{0.2109} & \textbf{79.05} \\
        \bottomrule
    \end{tabular}
    \label{tab:oos_r2}
\end{table}

\subsection{Out-of-time performance evaluation}

Table~\ref{tab:tab4} reports performance under the out-of-time protocol, where models are trained on historical data and evaluated on future test years. This walk-forward design mirrors realistic deployment conditions requiring models to generalize across shifting market regimes.

The R-GCN achieves a superior average $R^2$ of 76.79\%, surpassing XGBoost (68.88\%) and Random Forest (66.28\%) by 7.91 and 10.51 percentage points (pp), respectively. Its average MSE of 0.1511 represents a 24.1\% reduction relative to XGBoost. Notably, the R-GCN's improvement margin over tabular baselines is substantially larger in the OOT setting than in the OOS context (7.91pp vs.\ 1.42pp), suggesting that relational graph structures are particularly robust for temporal extrapolation.

The R-GCN outperforms benchmarks in five of the six temporal splits, reaching its peak $R^2$ in 2021 (89.03\%). Its most significant advantage occurs in 2018, where its $R^2$ of 80.04\% exceeds XGBoost’s 53.48\% by a margin of 26.56pp. Although XGBoost leads the 2019 cohort (72.97\% vs.\ 63.14\%), the R-GCN retains its superior performance during the 2020 COVID-19 market disruption ($R^2$: 51.77\%).

This enhanced generalization is attributed to the message-passing mechanism, which propagates pricing signals through stable entity-level relationships—including cedents, underwriters, and perils—that persist across market cycles. While individual bond features fluctuate across regimes, the relational topology provides a time-invariant learning signal that tabular models, operating on isolated feature vectors, cannot exploit.

\begin{table}[htbp]
    \centering
    \caption{OOT model performance across walk-forward splits}
    % \resizebox{\textwidth}{!}{ % Uncomment if the table is too wide
    \begin{tabular}{lccccccc}
        \toprule
        \multirow{2}{*}{\textbf{Train / Val Split}} & \multirow{2}{*}{\textbf{Test Year}} & \multicolumn{2}{c}{\textbf{Random Forest}} & \multicolumn{2}{c}{\textbf{XGBoost}} & \multicolumn{2}{c}{\textbf{R-GCN}} \\
        \cmidrule(lr){3-4} \cmidrule(lr){5-6} \cmidrule(lr){7-8}
        & & \textbf{MSE} & \textbf{$\mathbf{R^2}$ (\%)} & \textbf{MSE} & \textbf{$\mathbf{R^2}$ (\%)} & \textbf{MSE} & \textbf{$\mathbf{R^2}$ (\%)} \\
        \midrule
        1999--2014 | Val 2015 & 2016 & 0.1722 & 74.92 & 0.1205 & 82.45 & 0.0820 & 88.05 \\
        1999--2015 | Val 2016 & 2017 & 0.1578 & 73.39 & 0.1149 & 80.63 & 0.0668 & 88.73 \\
        1999--2016 | Val 2017 & 2018 & 0.2335 & 62.01 & 0.2860 & 53.48 & 0.1227 & 80.04 \\
        1999--2017 | Val 2018 & 2019 & 0.2295 & 69.13 & 0.2010 & 72.97 & 0.2741 & 63.14 \\
        1999--2018 | Val 2019 & 2020 & 0.3252 & 45.01 & 0.3237 & 45.27 & 0.2853 & 51.77 \\
        1999--2019 | Val 2020 & 2021 & 0.1851 & 73.23 & 0.1489 & 78.46 & 0.0758 & 89.03 \\
        \midrule
        \textbf{Average} & \textbf{--} & \textbf{0.2172} & \textbf{66.28} & \textbf{0.1992} & \textbf{68.88} & \textbf{0.1511} & \textbf{76.79} \\
        \bottomrule
    \end{tabular}
    \label{tab:tab4}
    % }
\end{table}

It is important to contextualize these results (i.e., OOS and OOT) within the existing literature. While other studies on similar datasets have reported high R\textsuperscript{2} values, those outcomes often rely on extensive, manual feature engineering, external data,  and data manipulation. In contrast, the performance lift demonstrated here is achieved by leveraging the graph representation itself to automatically extract value from the data's relational structure, without subjective, manual intervention. 

We validated the assumption that macroeconomic information is already embedded in catastrophe bond prices at issuance by introducing two market indicators: the BBB corporate credit spread and the Guy Carpenter Rate-on-Line (see Appendix~\ref{with_macros}, Tables~\ref{tab:tab11}--\ref{tab:tab12}). As summarized in our findings, these macroeconomic covariates produced asymmetric effects across model architectures. In the OOS setting, Random Forest and XGBoost improved slightly (+2.57\% and +0.47\%, respectively), while the R-GCN declined by 2\%. The out-of-time (OOT) setting revealed a sharper divide: while Random Forest maintained a 2.39\% gain, XGBoost deteriorated significantly ($-$4.34\%) and the R-GCN declined further ($-$2.51\%).

These divergent responses reflect architectural differences in covariate processing. Random Forest’s bagging and feature sub-sampling naturally regularize against uninformative inputs. Conversely, XGBoost’s sequential boosting aggressively fits residual patterns, yielding modest OOS gains but substantial OOT degradation as market regimes shift—a byproduct of the non-stationary nature of credit spread dynamics.

The R-GCN’s sensitivity arises from its message-passing mechanism. Because bonds issued within the same year share temporal edges, macroeconomic features—which are identical for the entire cohort—are reinforced through neighborhood aggregation. This amplification causes the model to overweight uniform temporal signals at the expense of primary pricing drivers like expected loss or maturity. Unlike tabular models that score observations independently, the graph structure transforms these covariates into a network-wide shift in learned representations.

In summary, the results show that tabular models, even when given the same initial contract information, may not adequately exploit the complex web of relationships within catastrophe bond data. The graph representation itself provides a foundational performance lift, and the subsequent inclusion of network-derived features offers a second, significant improvement, leading to a state-of-the-art predictive model.

\subsection{Feature importance and business intuition}

 A common critique of deep learning models is their perceived ``black box" nature, where the rationale behind predictions can be nontransparent. To address this limitation and provide clear business and economic intuition, we employ GNNExplainer \citep{A20} to interpret the predictions of our R-GCN model. GNNExplainer is a model-agnostic tool designed to identify the most influential components, both features and underlying graph structures, that contribute to a given prediction. This allows us to move beyond performance metrics and understand the key drivers of bond pricing at a granular level.

 The explainer works by identifying a compact subgraph and a small subset of node features that are maximally influential for the model's prediction. As framed by \cite{A20}, this task can be viewed as an optimization problem that maximizes the mutual information between the model's output and a distribution of possible subgraph structures.

 Using this tool, we analyze feature importance at two distinct levels: the node level, comprising the quantitative features of the bond contract and the topological features generated from the graph structure, and the edge level, representing the qualitative relationships between entities (e.g., a specific underwriter's connection to a certain peril type). This dual analysis enables a comprehensive understanding of not just what features are important, but also how the relationships between market participants drive the model's predictions.

\begin{figure}[H]
    \centering
    \includegraphics[scale= 0.8]{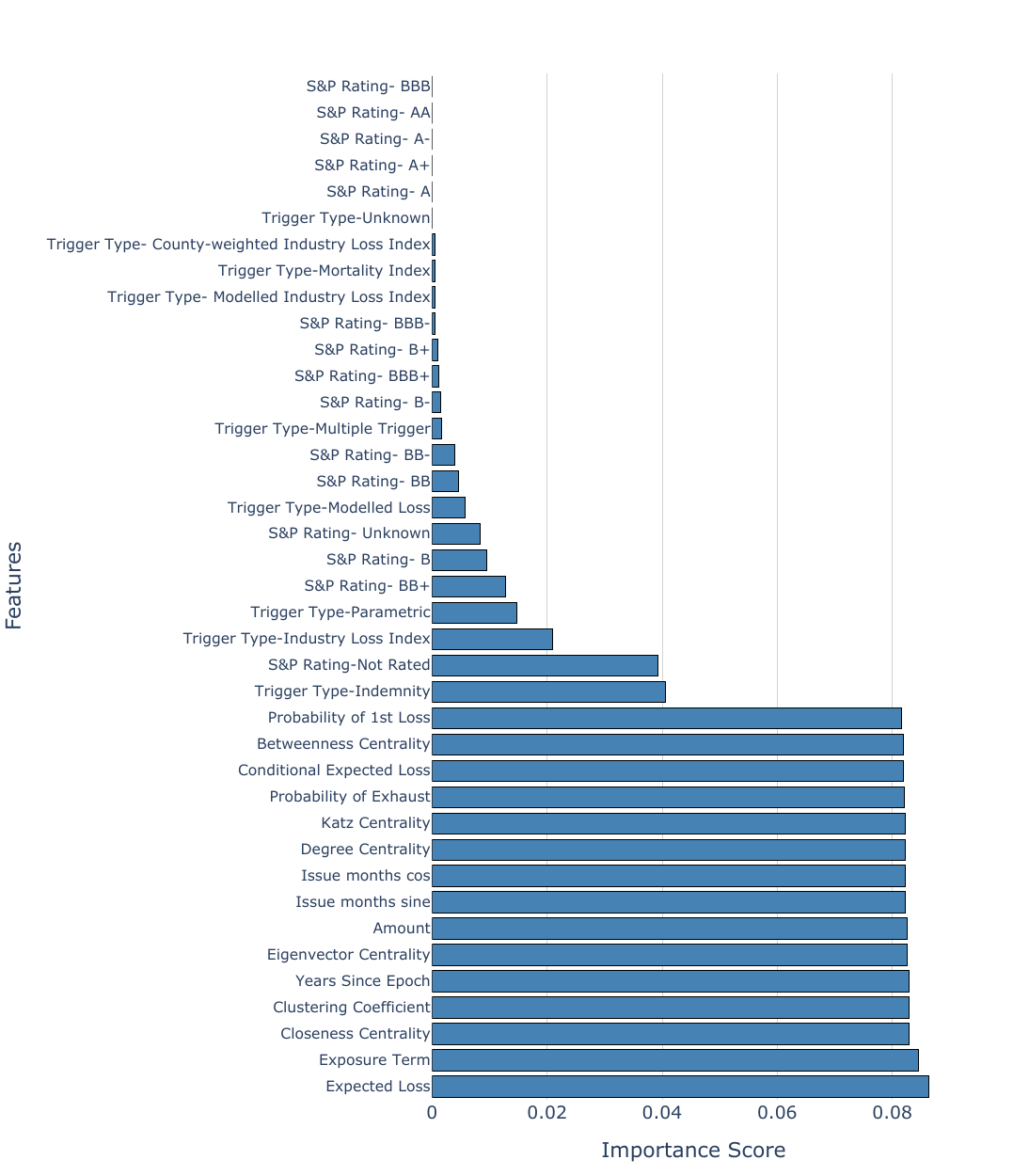}
    \caption{Ranked node feature importance.}
    \label{fig9}
\end{figure}

In Figure~\ref{fig9} we report the ranking of node features. The R-GCN’s top predictors for CAT bond premium – expected loss, probability of loss, conditional loss, bond size, term, timing, and network centrality measures – each correspond to meaningful drivers grounded in finance and insurance literature (see \cite{A30}, \cite{Braun2016}, \cite{A21}, \cite{Chen2024}). Expected loss and probability of first loss form the core risk pricing variables, long acknowledged as the primary determinants of CAT bond spreads \citep{Chen2024}.

\subsubsection{Core risk metrics}

\noindent \textbf{Expected Loss (EL):} As the modeled average loss, EL is the single most important driver of catastrophe bond spreads, a fact confirmed by numerous empirical studies \citep{Chen2024} and our own baseline models. Industry practitioners often quote prices as a ``multiple'' of EL, reflecting its role as the fundamental baseline for risk pricing \citep{A23}. The R-GCN's strong reliance on EL affirms its ability to capture the primary risk-return relationship.

\noindent \textbf{Probability of First Loss (PFL):} Represents the likelihood of a trigger event. PFL provides crucial information about the shape of the risk profile that EL alone does not capture. As first demonstrated by \cite{A30}, two bonds with the same EL can have different risk profiles based on the frequency of potential losses. For instance, investors may price a bond with frequent, small potential losses differently than one with a rare but severe potential loss. The model's focus on PFL aligns with industry practice, where spreads are directly correlated with the modeled probability of loss \citep{A22}.

\noindent \textbf{Conditional Expected Loss (CEL):} This is defined as the expected loss given a trigger event. CEL measures tail severity. Introduced by \cite{A30} to account for the fat-tailed nature of catastrophe risk, a higher CEL indicates that losses, when they occur, are more severe. The R-GCN's use of CEL as a key feature suggests it correctly prices in the additional risk premium that investors demand for exposure to extreme, high-severity outcomes.

\subsubsection{Contractual and market-based features} 

\noindent \textbf{Issue Amount (Deal Size):} The effect of a bond's principal amount is nuanced, often serving as a proxy for liquidity and prevailing market conditions. While academic findings are mixed, with some showing larger issues command lower spreads due to liquidity benefits \citep{Braun2016}, deal size also correlates with market cycles \citep{Chen2024}. For example, ``hard'' markets (when capital is scarce) often feature both higher spreads and smaller deal sizes. The R-GCN uses Issue Amount to capture these complex dynamics, linking deal size to the broader market environment.

\noindent \textbf{Exposure Term (Maturity):} The relationship between a bond's maturity and its premium is not linear. While a longer term increases the total time on risk, empirical studies have found a ``counter-intuitive relation'' where longer maturities do not always command higher annualized premiums \citep{Chatoro2023}. This can be due to factors like investors locking in yields or issuers taking advantage of favorable ``soft'' market conditions. The model's reliance on Exposure Term indicates it successfully captures these complex, non-linear pricing patterns that align with real-world observations.

\noindent \textbf{Issue Month (Seasonality):} Bond premiums often exhibit seasonality, driven by cyclical peril patterns (e.g., the North Atlantic hurricane season) and investor capital flows. Industry practice confirms that a bond's effective start date impacts its price, with higher premiums often demanded for bonds issued just before a peak risk season \citep{A24}. By identifying Issue Month as a top predictor, the R-GCN demonstrates its ability to learn these systematic seasonal trends.

\noindent \textbf{Issue Year (Market Cycle):} Catastrophe bond spreads are heavily influenced by market-wide ``reinsurance cycles'' of hardening and softening prices \citep{LaneMahul2008, Braun2016}. By using the Issue Year as a feature, the R-GCN effectively captures these macro-level temporal trends, learning that a bond's baseline premium depends significantly on the market environment at the time of its issuance. This aligns with other models that account for year-to-year market shifts \citep{A5}.

\subsubsection{Network topology features}\label{Section:topo}

As already discussed in Section \ref{sec:sec4.2.2}, a novel contribution of this work is the R-GCN's ability to exploit the graph structure of the CAT bond market, learning from the web of relationships connecting issuers, underwriters, and perils. The high predictive power of topological properties confirms that an entity's position and influence within this network carry significant pricing information. This finding provides a quantitative basis for the well-documented ``issuer effects'' and other dynamics in the CAT bond market \citep{Chatoro2023}.

\noindent \textbf{Closeness Centrality: A proxy for reputation and experience}

Closeness centrality measures how easily a node can reach all other nodes in the network. An entity with high closeness, such as a major issuer, is well-connected through many short paths, placing it at the ``center'' of market activity.

The R-GCN's reliance on this feature suggests it captures the critical role of issuer reputation and investor familiarity. Large, frequent issuers (e.g., USAA, Swiss Re, see Figure~\ref{fig11}) are central nodes in the market network. Their high centrality serves as a proxy for their experience, transparency, and investor trust. This aligns with empirical evidence that issuer identity alone explains a substantial portion of spread variation—approximately 26\% according to \cite{Chatoro2023}.

Essentially, the model learns that two bonds with identical risk metrics can price differently based on their sponsor. A well-known, central issuer may achieve a lower spread because investors are comfortable with their track record. Conversely, a new or infrequent issuer—a peripheral node in the network—may need to pay an ``unknown sponsor premium.'' This also explains pricing differentials observed for sponsors like Swiss Re, whose central role as an arranger corresponds to consistent pricing patterns learned by the model \citep{Chen2024}. By identifying closeness centrality as a key predictor, the R-GCN quantifies the very real ``issuer effect'' that practitioners have long observed. \\

\noindent \textbf{Betweenness Centrality: A proxy for brokerage and risk concentration}

Betweenness centrality identifies nodes that act as critical bridges or connectors, frequently lying on the shortest paths between other nodes. The importance of this feature reveals two other key market dynamics: the influence of intermediaries and the impact of peril concentration. 

For underwriters, high betweenness signifies a key market-maker. A central investment bank that structures many deals can leverage its distribution power to attract broad investor interest, potentially tightening spreads. The model learns that the ``broker'' matters. 

For perils, betweenness highlights risk concentration. A common peril like ``Florida Hurricane'' links many otherwise disconnected issuers and investors, giving it high betweenness. The model learns that investors, likely already holding this risk, demand a higher premium for such concentrated exposure. Conversely, a rare peril in an uncommon region acts as a diversifying asset. Its low betweenness corresponds to a lower risk premium, as investors value its diversification benefits (see example \cite{Chen2024}).

Thus, betweenness centrality allows the model to understand the value of both brokerage power and risk diversification—factors that are difficult to capture in traditional, non-network models.

\noindent \textbf{Eigenvector Centrality: core influence and systemic importance}

Eigenvector centrality identifies a node's influence based on the importance of its neighbors \citep{Chen2014}. A high score signifies a ``core player"—an entity connected to other highly connected entities. In the CAT bond market, this metric points to systemically important sponsors or perils. As seen in insurance networks, such centrality is a key contributor to systemic risk exposure \citep{Alves2015}.

Consequently, a bond linked to a pivotal node may embed a systemic risk premium, as investors demand compensation for risks that could trigger market-wide losses. Conversely, a central, reputable issuer might secure better pricing due to market familiarity, an effect that helps explain why issuer identity can account for ~26\% of price variation \citep{Chatoro2023}. The model's use of this feature confirms it can identify these core market players and their impact on pricing.

\noindent \textbf{Degree Centrality: direct connections and market integration}

Degree centrality is the simplest measure of connectivity, counting a node's direct links. In this context, it reflects an entity's level of direct market participation—for example, the number of bonds a sponsor has issued. A high degree often signals a strong reputation and a diversified risk-transfer strategy, which can lead to more favorable pricing due to investor familiarity and competition \citep{Chen2014}.

However, this effect has its limits. While a higher degree can reflect diversification (lowering spreads), extremely high connectivity can introduce contagion risk \citep{Gandica2020}. The R-GCN's reliance on this feature indicates it captures this dual effect, balancing the benefits of market integration against the risks of over-concentration.

\noindent \textbf{Clustering Coefficient: localized connectivity and risk concentration}

The clustering coefficient measures the interconnectedness of a node’s immediate neighbors. A high coefficient indicates that a node is part of a tightly-knit ``clique,'' signifying localized market segmentation and risk concentration. For example, a cluster could consist of several bonds covering the same ``peak peril" held by an overlapping group of investors.

Financial network research shows that while high clustering facilitates rapid information flow, it also magnifies contagion. For CAT bonds, this means an adverse event can easily impact all members of a correlated cluster. Investors, therefore, demand higher spreads to compensate for this lack of diversification, a phenomenon observed in the higher risk premia for peak perils \citep{A23}. The model's use of this feature shows it can identify and price these pockets of concentrated risk.

\noindent \textbf{Katz Centrality: broad connectedness and indirect influence}

Katz centrality generalizes eigenvector centrality by better accounting for influence propagated through long chains of connections \citep{Gandica2020}. A node with high Katz centrality has a broad, indirect reach across the network, capable of creating ``ripple effects.''

In financial networks, this metric is strongly related to systemic risk and an entity's potential to trigger cascades \citep{Glasser2015}. In the CAT bond market, this means a peril or sponsor might be indirectly linked to numerous portfolios, even if its direct connections are modest. A bond associated with such a node carries a premium for this wider contagion potential. By identifying Katz centrality as a key predictor, the R-GCN demonstrates its sophisticated ability to look beyond immediate connections and price the risks associated with an entity's total influence on the entire market ecosystem (see Section~\ref{sec:section3.3}).

Finally, we translate the model's learned patterns into practical business application by examining the importance of the entities involved in each transaction. This analysis provides powerful, data-driven insights into how specific relationships and market players drive catastrophe bond pricing.

\begin{figure}[H]
    \centering
    \includegraphics[scale= 0.8]{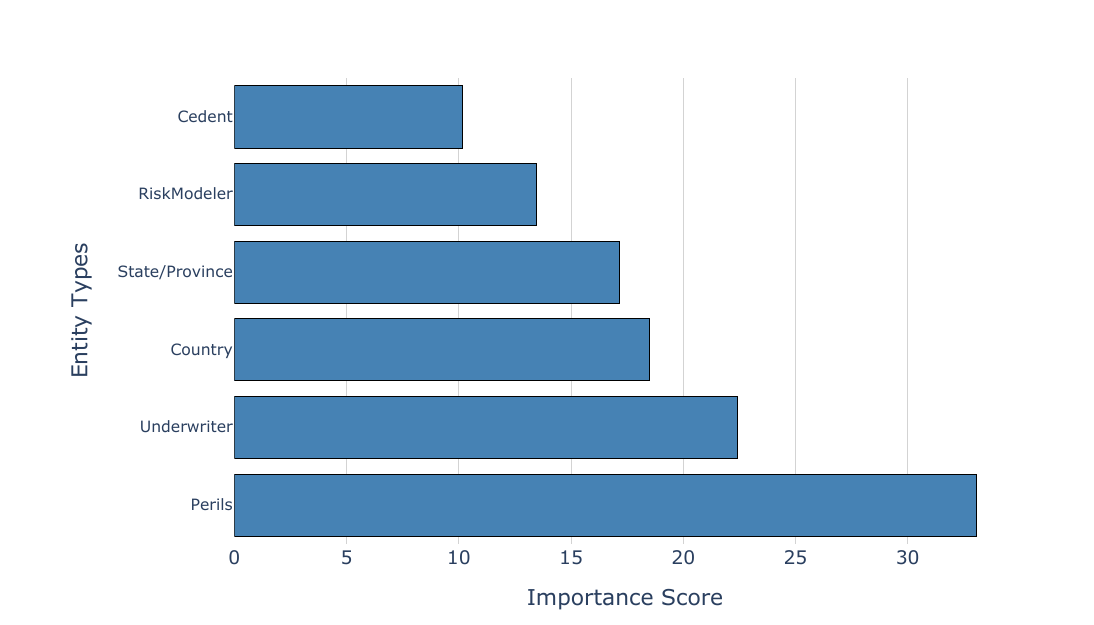}
    \caption{Ranking of group level edge importance by entity type}
    \label{fig10}
\end{figure}

\begin{figure}[H]
    \centering
    \includegraphics[scale= 0.8]{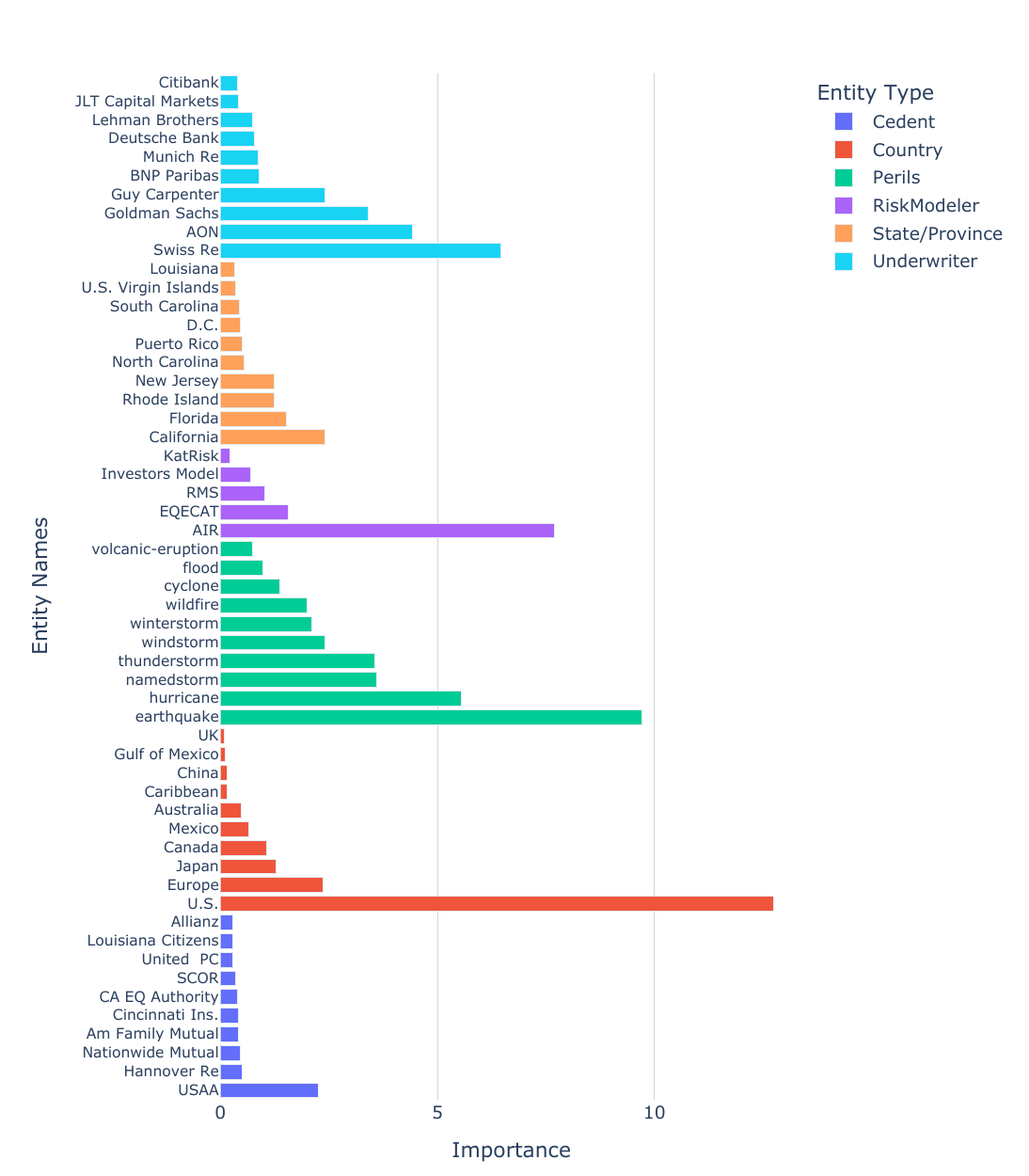}
    \caption{Ranking the edge importance of the top 10 entities by entity type}
    \label{fig11}
\end{figure}

First, we analyze the relative importance of each entity type (see Figure~\ref{fig10}). The results reveal a clear hierarchy of influence. With the highest importance score, the Perils category confirms the fundamental principle that the nature of the risk being transferred is the primary determinant of spread variability. Following perils, the Underwriter category ranks as the second most critical factor, validating our earlier findings on network effects where the reputation and market access of the deal's structurer significantly impact pricing. Finally, the high importance of Country and State/Province underscores that geographic location, tied directly to peril concentration and investor exposure, is a top-tier consideration.

To provide more granular and actionable insights, we then identify the specific entities within each category that most influence pricing. This fine-grained view, presented in Figure~\ref{fig11}, identifies the key market movers. Among Cedents, frequent and established issuers like USAA and Hannover Re emerge as highly important, reflecting their status as market benchmarks. Foundational Perils such as earthquake and hurricane are confirmed as key drivers. Notably, the analysis highlights the dominant influence of AIR as the most critical Risk Modeler, suggesting its methodologies significantly affect investor perception and, therefore, pricing. In the Underwriter category, the prominence of major players like Swiss Re and AON confirms their central role in structuring the market. Finally, the analysis shows that  the United States is the geographic mainstay of most CAT transactions with the sates of California and Florida having the most exposure due the frequent occurrence of earthquakes and hurricane respectively.. 

This ability to move from abstract factors to specific, influential entities is a distinct advantage of our graph-based approach, providing data-driven evidence of who and what truly drives risk pricing in the catastrophe bond market.

\newpage
\section{Concluding remarks} \label{sec6}

This paper addresses the challenge of pricing catastrophe (CAT) bonds by introducing CATNet, a novel framework that applies a geometric deep learning architecture, the Relational Graph Convolutional Network (R-GCN), to CAT bond pricing in the primary market. By modeling the market as a graph, this approach demonstrates that its underlying network structure is a powerful and previously underutilized source of predictive information.

Our results revealed a significant advantage of this approach. We found that the CAT bond market exhibits the characteristics of a scale-free network, a structure dominated by a few highly connected ``hubs" and many sparsely connected entities. This architecture, while efficient, introduces systemic vulnerabilities, as shocks to these central nodes can propagate widely. CATNet's success stems directly from its ability to navigate this structure. CATNet significantly outperformed a strong Random Forest and XGBoost benchmarks. More importantly, interpretability analysis showed that these centrality measures are not statistical artifacts; they are quantitative proxies for the influence of these network hubs, effectively capturing long-held industry intuition about issuer reputation, underwriter influence, and peril concentration.

The findings suggest a new paradigm for pricing complex, relationship-heavy financial instruments, shifting the focus from manual feature engineering to learning directly from network structures. While the model demonstrated robust out-of-time performance, we acknowledge its sensitivity to data scarcity in certain market periods and regime changes. A promising avenue for future research is the application of dynamic temporal graph neural networks, which could explicitly model the market regime shifts that our analysis identified, potentially capturing how the market's structure and pricing dynamics evolve over time.

In conclusion, this research provides evidence that in the catastrophe bond market, connectivity is a key determinant of price. By applying our CATNet framework, we have shown that it is not only possible to achieve state-of-the-art prediction accuracy but also to gain a deeper, more quantitative understanding of the intricate, scale-free relationships that govern this primary CAT bond market.

\newpage
\bibliographystyle{}

\newpage
\appendix

\section{Appendix A}
\subsection{Data description }

\begin{table}[!ht]
\centering
\scriptsize % Reduce font size
\begin{tabularx}{\columnwidth}{lXll} % Use columnwidth for smaller width
\toprule
Categorical variables & Levels & Count & Percentage (\%) \\
\midrule
S\&P Rating & NR & 350 & 43.59 \\
 & BB+ & 127 & 15.82 \\
 & UNKNOWN & 76 & 9.46 \\
 & B & 63 & 7.85 \\
 & BB & 58 & 7.22 \\
 & BB- & 53 & 6.60 \\
 & B+ & 27 & 3.36 \\
 & B- & 22 & 2.74 \\
 & BBB- & 13 & 1.62 \\
 & BBB+ & 6 & 0.75 \\
 & A- & 4 & 0.50 \\
 & A & 1 & 0.12 \\
 & BBB & 1 & 0.12 \\
 & A+ & 1 & 0.12 \\
 & AA & 1 & 0.12 \\[5pt] % Add space between categories
TriggerType & Indemnity & 359 & 43.25 \\
 & Industry Loss Index & 224 & 26.99 \\
 & Parametric & 154 & 18.55 \\
 & Modelled Loss & 44 & 5.30 \\
 & Multiple Trigger & 31 & 3.73 \\
 & UNKNOWN & 6 & 0.72 \\
 & County-weighted Industry Loss Index & 5 & 0.60 \\
 & Mortality Index & 4 & 0.48 \\
 & Modelled Industry Loss Index & 3 & 0.36 \\[5pt] % Add space between categories
RiskModeler & AIR & 411 & 51.18 \\
 & UNKNOWN & 175 & 21.79 \\
 & RMS & 109 & 13.57 \\
 & EQECAT & 92 & 11.46 \\
 & KatRisk & 8 & 1.00 \\
 & Investors Model & 5 & 0.62 \\
 & Aon & 1 & 0.12 \\
 & Multiple & 1 & 0.12 \\
 & Towers Watson & 1 & 0.12 \\[5pt] % Add space between categories
Perils & earthquake & 556 & 29.32 \\
 & hurricane & 321 & 16.93 \\
 & namedstorm & 226 & 11.92 \\
 & windstorm & 194 & 10.23 \\
 & thunderstorm & 158 & 8.33 \\
 & winterstorm & 104 & 5.49 \\
 & wildfire & 97 & 5.12 \\
 & cyclone & 59 & 3.11 \\
 & flood & 43 & 2.27 \\
 & typhoon & 34 & 1.79 \\
 & meteorite-impact & 29 & 1.53 \\
 & volcanic-eruption & 29 & 1.53 \\
 & multi-peril & 14 & 0.74 \\
 & tornado & 6 & 0.32 \\
 & brushfire & 6 & 0.32 \\
 & hailstorm & 6 & 0.32 \\
 & UNKNOWN & 4 & 0.21 \\
 & atmospheric-peril & 4 & 0.21 \\
 & mortality & 3 & 0.16 \\
 & temperature & 2 & 0.11 \\
 & snowstorm & 1 & 0.05 \\
\bottomrule
\end{tabularx}
\caption{Descriptive statistics on the relational nature of CAT contracts}
\label{tab1}
\end{table}

\newpage
\begin{table}[H]
\centering
\scriptsize % Reduce font size
\begin{tabularx}{\columnwidth}{lXll} % Use columnwidth for smaller width
\toprule
Categorical variables & Levels & Count & Percentage (\%) \\
\midrule
Underwriter & Swiss Re & 341 & 25.70 \\
 & AON & 235 & 17.71 \\
 & Goldman Sachs & 206 & 15.52 \\
 & Guy Carpenter & 164 & 12.36 \\
 & Deutsche Bank & 65 & 4.90 \\
 & Willis Capital Markets & 49 & 3.69 \\
 & BNP Paribas & 47 & 3.54 \\
 & Munich Re & 45 & 3.39 \\
 & Citibank & 36 & 2.71 \\
 & Lehman Brothers & 29 & 2.19 \\
 & Merrill Lynch & 21 & 1.58 \\
 & NT & 14 & 1.06 \\
 & BoA & 10 & 0.75 \\
 & Tiger Capital Markets & 7 & 0.53 \\
 & MMC Securities & 7 & 0.53 \\
 & Rewire Securities & 7 & 0.53 \\
 & AIG & 6 & 0.45 \\
 & American Re & 6 & 0.45 \\
 & ABN AMRO & 5 & 0.38 \\
 & SDD & 4 & 0.30 \\
 & BP & 4 & 0.30 \\
 & JLT Capital Markets & 3 & 0.23 \\
 & JP Morgan Chase & 3 & 0.23 \\
 & Towers Watson Capital Markets & 2 & 0.15 \\
 & LCM & 2 & 0.15 \\
 & Morgan Stanley & 2 & 0.15 \\
 & UBS & 2 & 0.15 \\
 & E.W. Blanch & 2 & 0.15 \\
 & Hanover Re & 1 & 0.08 \\
 & CDC IXIS & 1 & 0.08 \\
 & UNKNOWN & 1 & 0.08 \\[5pt] % Add space between categories
Country & U.S. & 670 & 54.74 \\
 & Europe & 184 & 15.03 \\
 & Japan & 133 & 10.87 \\
 & Canada & 80 & 6.54 \\
 & Mexico & 29 & 2.37 \\
 & Australia & 15 & 1.23 \\
 & UK & 15 & 1.23 \\
 & France & 14 & 1.14 \\
 & Belgium & 8 & 0.65 \\
 & Germany & 8 & 0.65 \\
 & Netherlands & 8 & 0.65 \\
 & Ireland & 7 & 0.57 \\
 & UNKNOWN & 7 & 0.57 \\
 & Caribbean & 7 & 0.57 \\
 & Denmark & 7 & 0.57 \\
 & Luxembourg & 6 & 0.49 \\
 & Italy & 4 & 0.33 \\
 & Turkey & 3 & 0.25 \\
 & Switzerland & 2 & 0.16 \\
 & Norway & 2 & 0.16 \\
 & Sweden & 2 & 0.16 \\
 & Mediterranean & 2 & 0.16 \\
 & Philippines & 2 & 0.16 \\
 & Gulf of Mexico & 1 & 0.08 \\
 & Portugal & 1 & 0.08 \\
 & Spain & 1 & 0.08 \\
 & Madrid & 1 & 0.08 \\
 & Taiwan & 1 & 0.08 \\
 & China & 1 & 0.08 \\
 & Chile & 1 & 0.08 \\
 & Colombia & 1 & 0.08 \\
 & Peru & 1 & 0.08 \\
\bottomrule
\end{tabularx}
\caption{Descriptive statistics on underwriters and country representation}
\label{tab2}
\end{table}

\newpage

%###############

\begin{table}[H]
\centering
\scriptsize
\begin{tabularx}{\columnwidth}{lXll} 
\toprule
Categorical variables & Levels & Count & Percentage (\%) \\
\midrule
Cedent & Swiss Re & 178 & 21.92 \\
&USAA & 75 & 9.24 \\
&Munich Re & 30 & 3.69 \\
&Hannover Re & 27 & 3.33 \\
&SCOR & 20 & 2.46 \\
&Nationwide Mutual & 19 & 2.34 \\
&Everest Re & 19 & 2.34 \\
&CA EQ Authority & 18 & 2.22 \\
&Allianz & 18 & 2.22 \\
&XL Bermuda & 15 & 1.85 \\
&Zenkyoren Ins. & 15 & 1.85 \\
&State Farm & 15 & 1.85\\
&IBRD & 14 & 1.72 \\
&Allstate & 12 & 1.48 \\
&Chubb Group & 10 & 1.23 \\
&Assurant & 9 & 1.11 \\
&Tokio Marine & 8 & 0.99 \\
&AIG & 8 & 0.99 \\
&National Union & 8 & 0.99 \\
&Heritage PC & 8 & 0.99 \\
&Liberty Mutual & 8 & 0.99 \\
&Travellers & 7 & 0.86 \\
&Catlin Ins & 7 & 0.86 \\
&Safepoint Ins Co & 7 & 0.86 \\
&Citizen's Property Ins. & 7 & 0.86 \\
&Hartford & 7 & 0.86 \\
&Am Strategic Ins & 6 & 0.74 \\
&FEMA & 6 & 0.74 \\
&Louisiana Citizens & 6 & 0.74 \\
&Sompo Nipponkoa & 5 & 0.62 \\
&Argo Re & 5 & 0.62 \\
&AXIS Re & 5 & 0.62 \\
&Mitsui Sumitomo & 5 & 0.62 \\
&United PC & 5 & 0.62 \\
&Palomar Specialty Ins. & 5 & 0.62 \\
&Avatar PC & 5 & 0.62 \\
&AXA Global & 4 & 0.49 \\
&Am Re & 4 & 0.49 \\
&Glacier Re & 4 & 0.49 \\
&Arrow Re & 4 & 0.49 \\
&Fidelis Ins. & 4 & 0.49 \\
&CIG Re & 4 & 0.49 \\
&Am Integrity & 4 & 0.49 \\
&Nephila Capital Ltd. & 4 & 0.49 \\
&PXRE & 4 & 0.49 \\
&Bayview Opp Fd & 4 & 0.49 \\
&CA St Comp Ins Fd & 3 & 0.37 \\
&UnipolSai Ass SpA & 3 & 0.37 \\
&Amlin AG & 3 & 0.37 \\
&Flagstone & 3 & 0.37 \\
&FONDEN & 3 & 0.37 \\
&Great American Ins. & 3 & 0.37 \\
&Cincinnati Ins. & 3 & 0.37 \\
&Renaissance Re & 3 & 0.37 \\
&OCIL & 3 & 0.37 \\
&XL Insurance & 3 & 0.37 \\
&Validus Re & 3 & 0.37 \\
&NC Ins. Underwriting Assn. & 3 & 0.37 \\
&CEA & 3 & 0.37 \\
&Zurich & 3 & 0.37 \\
&Brit Ins. Holdings & 3 & 0.37 \\
&Castle Key Ins & 3 & 0.37 \\
&MMM IARD SA & 3 & 0.37 \\
&Tokio Millenium Re & 3 & 0.37 \\
\bottomrule
\end{tabularx}
\caption{Descriptive statistics on cedent representation (Part 1)}
\label{tab3}
\end{table}

\newpage
\begin{table}[H]
\centering
\scriptsize
\begin{tabularx}{\columnwidth}{lXll} 
\toprule
Categorical variables & Levels & Count & Percentage (\%) \\
\midrule
Cedent &Transatlantic Re & 3 & 0.37 \\
&Endurance Sp. Ltd. & 3 & 0.37 \\
&Achmea Re & 2 & 0.25 \\
&American Family Ins & 2 & 025 \\
&Alphabet & 2 & 0.25 \\
&First Prot Ins & 2 & 0.25 \\
&ICAT Syndicate & 2 & 0.25 \\
&Convex Re & 2 & 0.25 \\
&TWIA & 2 & 0.25 \\
&DaVinci & 2 & 0.25 \\
&USFG & 2 & 0.25 \\
&Gerling & 2 & 0.25 \\
&Vesta wildfire Ins. & 2 & 025 \\
&Montpelier Re & 2 & 0.25 \\
&FM Global & 2 & 0.25 \\
&AGF & 2 & 0.25 \\
&Koch & 2 & 0.25\\
&Federal Ins. Co. & 2 & 0.25 \\
&Flagstone Re Ltd & 2 & 0.25 \\
&Oriental Land & 2 & 0.25 \\
&Nissay Dowa & 2 & 0.25 \\
&Flagstone Re & 2 & 0.25 \\
&Am Family Mutual & 2 & 0.25 \\
&Natixis SA & 2 & 0.25 \\
&Sorema & 2 & 0.25 \\
&Kemper& 2 & 0.25 \\
&Turkish Cat Ins Pool & 2 & 0.25 \\
&American Coastal Ins & 2 & 0.25 \\
&First Mutual Trans & 2 & 0.25 \\
&Lehman Re & 2 & 0.25 \\
&Sempra En & 1 & 0.12 \\
&Hiscox Syndicate & 1 & 0.12 \\
&Vivendi & 1 & 0.12 \\
&Oak Tree Assur & 1 & 0.12 \\
&Amer Modern Ins & 1 & 0.12 \\
&Markel Bermuda & 1 & 0.12 \\
&FMTA & 1 & 0.12 \\
&Electricite de France & 1 & 0.12 \\
&Allied World & 1 & 0.12 \\
&Hamilton Re & 1 & 0.12 \\
&Brit Syndicates & 1 & 0.12 \\
&Universal PC & 1 & 0.12 \\
&Central Re Corp. & 1 & 0.12 \\
&Aura Re & 1 & 0.12 \\
&Texas Windst Ins. Assn & 1 & 0.12 \\
&Aspen Bermuda Ltd. & 1 & 0.12 \\
&NJ Manuf Ins & 1 & 0.12 \\
&Florida Muni Ins Tr & 1 & 0.12 \\
&Generali & 1 & 0.12 \\
&China PC & 1 & 0.12 \\
&Passenger Railroad Ins. & 1 & 0.12 \\
&Platinum & 1 & 0.12 \\
&Groupama & 1 & 0.12 \\
&Aspen Ins. Holdings & 1 & 0.12 \\
&Balboa Ins & 1 & 0.12 \\
&Dominion Resources & 1 & 0.12 \\
&Mass Property & 1 & 0.12 \\
&Assicurazioni Generali & 1 & 0.12 \\
&AmTrust Fin Svc & 1 & 0.12 \\
&Equator Re Ltd & 1 & 0.12 \\
&Converium & 1 & 0.12 \\
&GI Capital Ltd. & 1 & 0.12 \\
&Aioi Nissay Dowa & 1 & 0.12 \\
&Security First Ins. & 1 & 0.12 \\
\bottomrule
\end{tabularx}
\caption{Descriptive statistics on cedent representation (Part 2)}
\label{tab4}
\end{table}
%##############
\begin{table}[H]
\centering
\caption{Summary statistics of numerical variables}
\begin{tabular}{lrr}
\hline
\textbf{Variable} & \textbf{Mean} & \textbf{Standard Deviation} \\
\hline
Issue amount (M\$)                    & 134.34 & 115.46 \\
Expected excess return &  0.0533  & 0.0371\\
Spread premium to LIBOR       &  0.0758 &  0.0503\\
Expected loss               & 0.0003 & 0.0004 \\
Probability of 1st loss       & 0.0024 & 0.0084 \\
Probability of exhaust       & 0.0016 & 0.0056 \\
Conditional expected loss    & 0.0078 & 0.0319 \\
\hline
\end{tabular}
\label{tab5}
\end{table}

\section{Appendix B}

\subsection{Hyperparameters} \label{hyperparameters}

Hyperparameter optimization for the R-GCN model was performed using the Optuna framework, which employs a Bayesian optimization strategy to efficiently search the parameter space. The objective was to minimize validation loss while avoiding overfitting through early stopping. Table~\ref{tab:search_space} summarizes the search space, including both continuous and categorical hyperparameters. The ranges for continuous parameters were chosen to balance exploration of wide value intervals with focus on practical ranges identified in prior GNN literature, while categorical choices reflect common architectural and optimizer configurations for R-GCNs.  

\begin{table}[H]
\centering
\caption{Hyperparameter search space for R-GCN model optimization}
\label{tab:search_space}
\begin{tabular}{lll}
\hline
\textbf{Hyperparameter} & \textbf{Range / Options} & \textbf{Type} \\ \hline
Learning rate & $10^{-6}$ to $10^{-2}$ (log-uniform) & Continuous \\
Hidden units & \{16, 32, 64, 128, 256\} & Categorical \\
Dropout rate & 0.0 to 0.5 & Continuous \\
Optimizer & \{Adam, SGD\} & Categorical \\
Activation function & \{ReLU, LeakyReLU, ELU, GELU\} & Categorical \\
Number of R-GCN layers & 1 to 5 & Integer \\ \hline
\end{tabular}
\end{table}

\subsection{Ablation study with macroeconomic variables} \label{with_macros}

\begin{table}[H]
    \centering
    \caption{OOS model performance across folds (with macro variables)}
    \begin{tabular}{lcccccc}
        \toprule
        \multirow{2}{*}{\textbf{Fold}} & \multicolumn{2}{c}{\textbf{Random Forest}} & \multicolumn{2}{c}{\textbf{XGBoost}} & \multicolumn{2}{c}{\textbf{R-GCN}} \\
        \cmidrule(lr){2-3} \cmidrule(lr){4-5} \cmidrule(lr){6-7}
        & \textbf{MSE} & \textbf{$\mathbf{R^2}$ (\%)} & \textbf{MSE} & \textbf{$\mathbf{R^2}$ (\%)} & \textbf{MSE} & \textbf{$\mathbf{R^2}$ (\%)} \\
        \midrule
        1 & 0.4021 & 64.03 & 0.3377 & 69.79 & 0.2813 & 74.83 \\
        2 & 0.1685 & 79.68 & 0.1983 & 76.08 & 0.1423 & 82.84 \\
        3 & 0.2879 & 75.33 & 0.2423 & 79.24 & 0.3394 & 70.92 \\
        4 & 0.2040 & 75.75 & 0.2302 & 72.63 & 0.2667 & 68.30 \\
        5 & 0.1260 & 79.75 & 0.1219 & 80.41 & 0.1478 & 76.25 \\
        6 & 0.4114 & 70.41 & 0.2479 & 82.18 & 0.2546 & 81.69 \\
        7 & 0.2039 & 82.59 & 0.2085 & 82.19 & 0.1911 & 83.68 \\
        8 & 0.3037 & 72.04 & 0.2939 & 72.94 & 0.3030 & 72.11 \\
        9 & 0.1672 & 77.98 & 0.1377 & 81.86 & 0.1638 & 78.43 \\
        10 & 0.2324 & 79.05 & 0.1808 & 83.70 & 0.2065 & 81.38 \\
        \midrule
        \textbf{Average} & \textbf{0.2507} & \textbf{75.66} & \textbf{0.2199} & \textbf{78.10} & \textbf{0.2296} & \textbf{77.04} \\
        \bottomrule
    \end{tabular}
    \label{tab:tab11}
\end{table}

\begin{table}[htbp]
    \centering
    \caption{OOT model performance across walk-forward splits (with macro variables) }
    % \resizebox{\textwidth}{!}{ % Uncomment if the table is too wide
    \begin{tabular}{lccccccc}
        \toprule
        \multirow{2}{*}{\textbf{Train / Val Split}} & \multirow{2}{*}{\textbf{Test Year}} & \multicolumn{2}{c}{\textbf{Random Forest}} & \multicolumn{2}{c}{\textbf{XGBoost}} & \multicolumn{2}{c}{\textbf{R-GCN}} \\
        \cmidrule(lr){3-4} \cmidrule(lr){5-6} \cmidrule(lr){7-8}
        & & \textbf{MSE} & \textbf{$\mathbf{R^2}$ (\%)} & \textbf{MSE} & \textbf{$\mathbf{R^2}$ (\%)} & \textbf{MSE} & \textbf{$\mathbf{R^2}$ (\%)} \\
        \midrule
        1999--2014 | Val 2015 & 2016 & 0.2318 & 66.23 & 0.1994 & 70.95 & 0.2341 & 65.90 \\
        1999--2015 | Val 2016 & 2017 & 0.1299 & 78.09 & 0.1027 & 82.68 & 0.0526 & 91.12 \\
        1999--2016 | Val 2017 & 2018 & 0.3056 & 51.61 & 0.2662 & 57.86 & 0.1505 & 76.17 \\
        1999--2017 | Val 2018 & 2019 & 0.2411 & 67.57 & 0.2296 & 69.12 & 0.2869 & 61.41 \\
        1999--2018 | Val 2019 & 2020 & 0.3212 & 45.61 & 0.2905 & 50.81 & 0.2498 & 57.70 \\
        1999--2019 | Val 2020 & 2021 & 0.1509 & 78.14 & 0.1338 & 80.62 & 0.0458 & 93.36 \\
        \midrule
        \textbf{Average} & \textbf{--} & \textbf{0.2301} & \textbf{64.54} & \textbf{0.2037} & \textbf{68.67} & \textbf{0.1700} & \textbf{74.28} \\
        \bottomrule
    \end{tabular}
    \label{tab:tab12}
    % }
\end{table}

\section{Appendix C}
\subsection{Network topology}\label{topo}

Following \cite{BarabasiPosfai2016}, we briefly review the basic concepts of a graph network discussed in section \ref{sec:section3.3}. \\

\noindent \textbf{Definition} (Cluster coefficient). For a graph $G= (\mathcal{V}(G), \mathcal{E}(G))$ of size $N$,  the local cluster coefficient of a given node $i\in V(G)$ with degree $k_i = d_i^{(G)}$ is given by: 

\begin{equation}
    C_i = \frac{1}{k_i (k_i -1)} \sum_{j, k \in \mathcal{V}(G)}\mathbb{I}_{  \{ij, jk, ik \in E(G)\} } = \frac{2L_i}{k_i (k_i -1)}
\end{equation}
where $L_i$ stands for the total number of links present in the neighbors of node $i$ (or equivalently the number of triangles that node $i$ forms with the two of its neighbors).  This quantity measures the fraction of the node's neighbors that are neighbors of each other, which takes its values between $0$ and $1$ with higher values indicating the more likely nodes in the neighborhood of that given node are connected\footnote{A statistical interpretation of relation (6) is in this way: Recall that the number of links for a network of size $N$ varies between $0$ and $\frac{N(N-1)}{2}$. Given that the degree of a specific node is $k$, it implies that there are $k$ nodes around that given node, resulting in $\frac{k(k - 1)}{2}$ possible links. Consequently, if there are $L$ links between neighbors of that given node, the probability that two neighbors of that given node are connected is given by
\begin{equation}
    \frac{L}{k(k -1)/2} = \frac{2L}{k (k -1)}
\end{equation}
}.\\

\noindent Another interesting quantity is the average clustering coefficient which can be seen as the probability that the two neighbors of a randomly selected node link together: 
\begin{equation}
   \langle  C  \rangle = \frac{1}{N} \sum_{i=1}^{N}C_i
\end{equation}
Finally, the global clustering coefficient which measures the total number of closed triangles over the graph, denoted by $C_{\Delta}$, is defined as follows (see, e.g., \cite{VanDerHofstad2024}): 
\begin{equation}
    C_{\Delta} = \frac{\sum_{1\leq i, j, k \leq N} \mathbb{I}_{ \{ ij, jk, ik \in \mathcal{E}(G)\} } }{\sum_{1\leq i, j, k \leq N} \mathbb{I}_{  \{ ij, jk \in \mathcal{E}(G)  \} }}=
    \frac{6 \sum_{1\leq i< j< k \leq N} \mathbb{I}_{ \{ ij, jk, ik \in \mathcal{E}(G)\} }   }{ 
  2 \sum_{1\leq i, j, k \leq N: i<k} \mathbb{I}_{  \{ ij, jk \in \mathcal{E}(G)  \} }   } = \frac{3\times \textit{Total number of triangles}}{\textit{Total number of triples}}
\end{equation}
%In the given formula, the denominator represents the total number of wedges in the graph, while the numerator specifies the total number of triangles. A wedge, or a triple, constitutes a subgraph with three vertices and edges connecting them. Notably, a wedge allows for the possibility of disconnectivity between nodes, representing a more general configuration. A triangle, on the other hand, specifically denotes a closed wedge where all three vertices are fully connected\footnote{The consideration of triangles in this context arises from the situation where the two neighbors of a given node establish a link between themselves, resulting in the formation of a triangle that includes the target node.}. It's essential to recognize that while every triangle is a wedge, not all wedges are triangles. The factor of 2 in the denominator accounts for the fact that, in an undirected graph, the edges $ij$ and $jk$ are equivalent to $ji$ and $kj$. This correction ensures that each unique wedge is counted only once. Furthermore, the factor of 6 arises from considering the permutations of the three vertices within each triangle. In an undirected graph, each triangle can be formed in six different ways by permuting its three vertices. This correction addresses the overcounting of triangles and ensures that each unique triangle is accounted for precisely once.\\

\noindent We continue with a quantity that measures the relationship between degrees of nodes. Within a network, nodes with high degrees might prefer establishing connections with either high-degree nodes or those with lower degrees. Based on the tendency of high-degree and low-degree nodes to form links with each other, networks can be categorized into three distinct types: Neutral, Assortative, and Disassortative. Neutral networks are those whose nodes are randomly linked, whereas in an assortative network, nodes with comparable degrees tend to connect to each other (i.e., small-degrees with small-degrees and hubs with hubs). Conversely, in a disassortative network, high-degree nodes are inclined to connect with small-degree nodes. This metric is referred to as the degree correlation which can be represented and quantified by using the so-called degree correlation matrix and degree correlation function, respectively. \\

\textbf{Definition} (Centrality measures). Several centrality measures can be taken to evaluate the importance of a node: Degree Centrality, Closeness Centrality, Betweeness Centrality, and eigenvector centrality.  The degree centrality of a given node is simply defined to be the node's degree. Nodes with higher degrees are deemed more important, as they are connected to a larger number of other nodes. For a given node $i$, the corresponding degree centrality is given by:
\begin{eqnarray}
    \text{DC}(i) = d_i^{(G)}
\end{eqnarray}
The closeness centrality measures how close a given node is to most other nodes. Nodes with higher closeness tend to be central in the network. The closeness centrality is defined as follows:
\begin{equation}
    \text{CC}(i) = \frac{1}{1/N\sum_{j \in \mathcal{V}(G)} dis_G(i, j)}
\end{equation}
where the denominator of the above fraction denotes the average length of the shortest path between the target node and all other nodes in the graph. If a node $j$ never reaches node $i$, for which $dis_G(i, j)= \infty$, $ \text{CC}(i) $ is zero. To address this limitation, a corrected version called harmonic closeness is introduced. This variation prevents the centrality measure from becoming zero while still accounting for disconnected nodes. 
\begin{eqnarray}
    \text{CC}(i) = \sum_{j \in \mathcal{V}(G)} \frac{1}{dis_G(i, j)}
\end{eqnarray}
The betweenness centrality quantifies the number of times that a given node lies between the shortest path of all other nodes, expressed as follows: 
\begin{eqnarray}
    \text{BC}(i) =  \sum_{1\leq j < k\leq N}  \frac{N_{jk}^{i}}{N_{jk}^{(G)}}  
\end{eqnarray}
where $N_{jk}^{(G)}$ is the total number of shortest path between nodes $j$ and $k$, and $N_{jk}^{i}$ is the total number of shortest path between node $j$ and $k$ that contains node $i$. Nodes with high betweenness centrality serve as a bridge that connects different parts of the graph to each other.\\

The eigenvector centrality of node $u$, denoted by $\text{EV}_u$, representing the importance of a node's neighbors, is derived using  the following recurrence relation: 
\begin{equation}
    \text{EV}_u = \frac{1}{\lambda} \sum_{v\in \mathcal{V}(G)} \mathbf{A}[u, v]
\end{equation}
where $\lambda$ is a constant. Apart from the above-mentioned node-level centrality measures, which basically shed light on the importance of individual nodes in a graph, there are sets of local and global overlap measures that quantify the relationship between neighbors of two nodes \citep{Hamilton2020}. One of the famous global overlap statistic is called Katz index, which is given by: 
\begin{equation}
   \text{Katz}[u] = \sum_{i = 0}^{+\infty} \beta^{i}\mathbf{A}[u, v] 
\end{equation}
where $\beta \in \mathbb{R}^{+}$ is a user-defined constant (decay factor) that determines how much influence indirect connections have. \\ 

\noindent Up to this point, our focus has been on deterministic graphs, where the number of edges is predetermined. Conversely, in a random graph, nodes are connected in a stochastic manner, and the number of edges becomes a random variable. The random graphs serve as a tool for mimicking the characteristics of real networks.\\

\noindent \textbf{Definition} (Random graph). A random graph $G=(\mathcal{V}(G), \mathcal{E}(G))$ of size $N$ and probability $p$, denoted as $G(N, p)$, is known to be a graph with deterministic vertex set $\mathcal{V}(G)$ and random vector $ (\mathbb{I}_{\{u, v\}})_{u, v\in \mathcal{V}(G)}$ for which each pair of nodes $u$ and $v$ are connected with equal probability $p$. \\

\noindent \textbf{Definition} (Scale-free graph). A scale-free graph $G=(\mathcal{V}(G), \mathcal{E}(G))$ of size $N$ satisfies the following properties:\\

\noindent - For some normalization $C$, the degree distribution of a scale-free network obeys a power-law distribution given by:
\begin{equation}
  p_k^{(G)}  = C k^{-\gamma}  
\end{equation}
where $\gamma$ is called the degree exponent.\\

\noindent The $n^{th}$ moment of the degree distribution is given by:
\begin{equation}
    \langle G^n\rangle = C \frac{k_{max}^{n-\gamma +1 } - k_{min}^{n-\gamma +1}}{n - \gamma +1}
\end{equation}
where the first moment is finite while moments of higher orders become infinite for large $N$, indicating a scale-free phenomenon. \\

\noindent - The probability of observing hubs in a scale-free network is higher than in a random network. The size of hubs increases polynomially with the size of the graph, i.e., $k_{max} = k_{min} N^{\frac{1}{\gamma -1}}$. \\

\noindent - A coexistence of widely varying degrees is observed, where numerous low-degree nodes are interconnected by a few number of highly connected hubs. \\

\noindent - For many scale-free networks, the degree exponent $\gamma$ typically falls within the range of $2$ to $3$, diverging notably from random networks where the parameter $\gamma$ tends to surpass $3$. \\

\noindent - The average distance in a scale-free network is smaller compared to its equivalent random network, indicating an ultra-small world phenomenon.

\begin{figure}[H]
    \centering
\includegraphics[width=\textwidth,height=0.9\textheight,keepaspectratio]{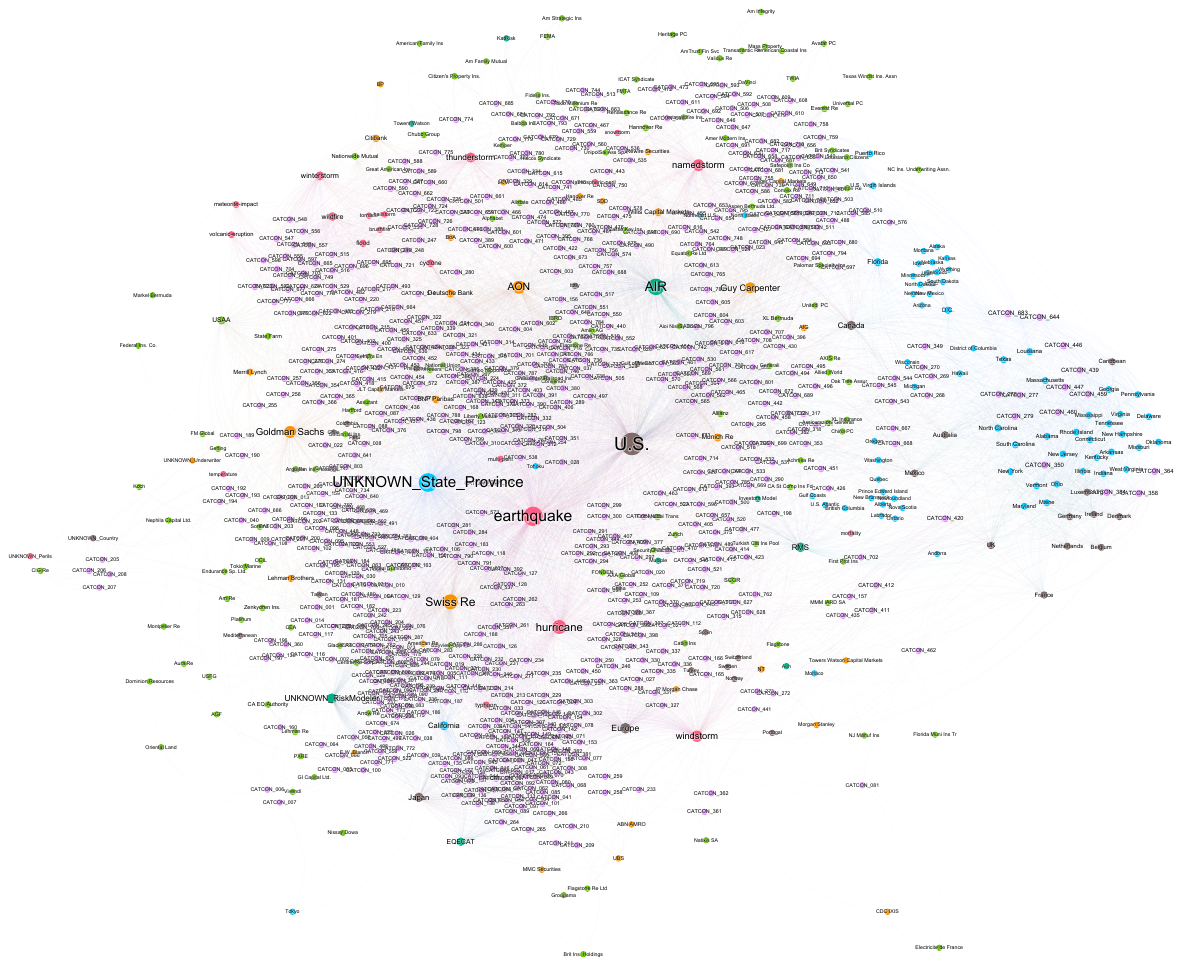}
    \caption{CAT bond network (1999–2021) visualization; larger nodes indicate higher degrees. Different edge colors signify unique relationships between the entities.}
    \label{fig3}
\end{figure}

\end{document}